\documentclass[twocolumn]{svjour3}          
\smartqed  
\usepackage{natbib}
\usepackage{graphicx}
\usepackage{xspace}
\usepackage{url}
\usepackage{eurosym}
\usepackage[polutonikogreek,english]{babel}

\graphicspath{{./Figures/}}
\newcommand{\uas}{\ensuremath{\mu\mbox{as}}\xspace}
\newcommand{\microns}{\ensuremath{\mu\mbox{m}}\xspace}
\newcommand{\meters}{\ensuremath{\mbox{m}}\xspace}
\newcommand{\AU}{\ensuremath{\mbox{AU}}\xspace}
\newcommand{\pc}{\ensuremath{\mbox{pc}}\xspace}

\newcommand{\Msun}{\ensuremath{\mbox{M}_\odot}\xspace}

\newcommand{\MEarth}{\ensuremath{\mbox{M}_\oplus}\xspace}
\newcommand{\SNR}{\ensuremath{\mbox{SNR}}\xspace}

\newcommand{\Nvisits}{\ensuremath{N_{\rm visits}}\xspace}

\newcommand{\Tvisits}{\ensuremath{T_{\rm visit}}\xspace}

\newcommand{\Mstar}{\ensuremath{M_\ast}\xspace}
\newcommand{\Mplanet}{\ensuremath{M_P}\xspace}

\urldef{\neaturl}\url{http://neat.obs.ujf-grenoble.fr}
\urldef{\neattargetsurl}\url{http://neat.obs.ujf-grenoble.fr/IMG/xls/Proposal_targets_total.xls}

\begin{document}

\title{High precision astrometry mission for the detection and characterization
  of nearby  habitable planetary systems with the Nearby Earth Astrometric 
  Telescope (NEAT)
}

\titlerunning{Nearby Earth Astrometric Telescope (NEAT)}

\author{
Fabien Malbet \and 
Alain L\'eger \and 
Michael Shao \and 
Renaud Goullioud \and
Pierre-Olivier Lagage \and
Anthony G.~A.\ Brown\and
Christophe Cara \and
Gilles Durand \and
Carlos Eiroa \and
Philippe Feautrier \and
Bj\"orn Jakobsson \and
Emmanuel Hinglais \and
Lisa Kaltenegger \and
Lucas Labadie \and
Anne-Marie Lagrange \and
Jacques Laskar \and
Ren\'e Liseau \and
Jonathan Lunine \and
Jes\'us Maldonado \and
Manuel Mercier \and
Christoph Mordasini \and
Didier Queloz \and
Andreas Quirrenbach \and
Alessandro Sozzetti \and
Wesley Traub \and
Olivier Absil \and
Yann Alibert \and
Alexandre Humberto Andrei \and
Fr\'ed\'eric Arenou \and
Charles Beichman \and
Alain Chelli \and
Charles S.\ Cockell \and
Gilles Duvert \and
Thierry Forveille \and 
Paulo J.V.~Garcia \and
David Hobbs \and
Alberto Krone-Martins \and
Helmut Lammer \and
Nad\`ege Meunier \and
Stefano Minardi \and
Andr\'e Moitinho de Almeida \and
Nicolas Rambaux \and
Sean Raymond \and
Huub J.~A.\ R\"ottgering \and 
Johannes Sahlmann \and
Peter A.\ Schuller \and
Damien S\'egransan \and
Franck Selsis \and
Jean Surdej \and 
Eva Villaver \and
Glenn J.\ White \and
Hans Zinnecker
}%

\authorrunning{Malbet, L\'eger, Shao, Goullioud, Lagage et al.} 

\institute{The complete affiliations are given at the end of the
  paper. The full list of members of the NEAT proposal is avialable at
  \texttt{http://neat.obs.ujf-grenoble.fr}. }

\date{Received: date / Accepted: date}

\maketitle

\begin{abstract}
  A complete census of planetary systems around a volume-limited
  sample of solar-type stars (FGK dwarfs) in the Solar neighborhood
  ($d\leq15\,pc)$ with uniform sensitivity down to Earth-mass planets
  within their Habitable Zones out to several AUs would be a major
  milestone in extrasolar planets astrophysics.  This fundamental goal
  can be achieved with a mission concept such as NEAT --- the Nearby
  Earth Astrometric Telescope.

  NEAT is designed to carry out space-borne extremely-high-preci\-sion
  astrometric measurements at the $0.05\,\mu$as ($1 \sigma$) accuracy
  level, sufficient to detect dynamical effects due to orbiting
  planets of mass even lower than Earth's around the nearest
  stars. Such a survey mission would provide the actual planetary
  masses and the full orbital geometry for all the components of the
  detected planetary systems down to the Earth-mass limit. The NEAT
  performance limits can be achieved by carrying out differential
  astrometry between the targets and a set of suitable reference stars
  in the field. The NEAT instrument design consists of an off-axis
  parabola single-mirror telescope (D = 1m), a detector with a large
  field of view located 40\,m away from the telescope and made of 8
  small movable CCDs located around a fixed central CCD, and an
  interferometric calibration system monitoring dynamical Young's
  fringes originating from metrology fibers located at the primary
  mirror. The mission profile is driven by the fact that the two main
  modules of the payload, the telescope and the focal plane, must be
  located 40\,m away leading to the choice of a formation flying
  option as the reference mission, and of a deployable boom option as
  an alternative choice. The proposed mission architecture relies on
  the use of two satellites, of about 700 kg each, operating at L2 for
  5 years, flying in formation and offering a capability of more than
  20,000 reconfigurations. The two satellites will be launched in a
  stacked configuration using a Soyuz ST launch vehicle.

  The NEAT primary science program will encompass an astrometric
  survey of our 200 closest F-, G- and K-type stellar neighbors, with an
  average of 50 visits each distributed over the nominal mission
  duration. The main survey operation will use approximately
  70\% of the mission lifetime.  The remaining ~30\% of NEAT observing
  time might be allocated, for example, to improve the
  characterization of the architecture of selected planetary systems
  around nearby targets of specific interest (low-mass stars, young
  stars, etc.) discovered by Gaia, ground-based high-precision
  radial-velocity surveys, and other programs. With its exquisite,
  surgical astrometric precision, NEAT holds the promise to provide the
  first thorough census for Earth-mass planets around stars in the
  immediate vicinity of our Sun.
\end{abstract}

\section{Introduction}
\label{sec:introduction}

Exoplanet research has grown explosively in the past decade, supported
by improvements in observational techniques that have led to
increasingly sensitive detection and characterization. Among many
results, we have learned that planets are common, but their physical
and orbital properties are much more diverse than originally
thought. 

A lasting challenge is the detection and characterization of planetary
systems consisting in a mixed cortege of telluric and giant planets,
with a special regard to telluric planets orbiting in the habitable zone
(HZ) of Sun-like stars. The accomplishment of this goal requires the
development of a new generation of facilities, due to the intrinsic
difficulty of detecting Earth-like planets with existing instruments.
The proposed NEAT mission has been designed to enter a
new phase in exoplanetary science by delivering an enhanced capability
of detecting small planets at and beyond 1\,AU.

Astrometry is probably the oldest branch of astronomy. Greeks
developed it and noticed that the position of most stars were stable
in the sky, but the few that were moving became known as
\emph{planets} (\textgreek{pl'anetes >ast'eres} = moving stars),
pointing to a major difference in their nature. Thanks to the precise
astrometric measurements of planet positions by Tycho Brahe in the
16th century, Johannes Kepler established that these objects were
orbiting the Sun on elliptical orbits, expanding the Copernican
revolution. After Hipparcos, Gaia will play an important role in
finding many systems with giant planets in our Galaxy. We want to
extend these revolutions with the NEAT mission, namely to discover and
characterize Earth-mass planets in Earth-like orbits around stars like
the Sun, by capturing infinitesimal displacements with unprecedented
accuracy.

In Sect.~\ref{sec:neat-science}, we present the science objectives of
NEAT, we describe the principle of the differential astrometry
technique and we give a list of potential targets. In
Sect.~\ref{sec:neat-concept}, after listing the technical challenges,
we present the instrumental concept. We explain how to reach the
performance and we give a summarized description of the payload, the
mission and the spacecraft. In Sect.~\ref{sec:discussion}, we discuss
both astrophysical and technical issues. Recommandations by the
community summarized in Sect.~\ref{sec:perspectives} is an incentive
to pursue the development of this mission in the future.

\section{NEAT Science}
\label{sec:neat-science}

\subsection{Science objectives}
\label{sec:science-cases}

The prime goal of NEAT is to detect and characterize planetary systems
orbiting bright stars in the solar neighborhood that have a planetary
architecture like that of our Solar System or an alternative planetary
system made of Earth mass planets. It will allow the detection around
nearby stars of planets equivalent to Venus, Earth, (Mars), Jupiter,
and Saturn, with orbits possibly similar to those in our Solar System.
It will permit to detect and characterize the orbits and the masses of
many alternate configurations, e.g.\ where the asteroid belt is
occupied by another Earth mass planet and no Jupiter. The NEAT mission
will answer the following questions:
\begin{itemize}
  \item What are the dynamical interactions between giant and
    telluric planets in a large variety of systems?
  \item What are the detailed processes involved in planet formation as
    revealed by their present configuration?
  \item What are the distributions of architectures of planetary systems
    in our neighborhood up to $\approx15$\,pc?
  \item What are the masses, and orbital parameters, of telluric planets that are
    candidates for future direct detection and spectroscopic characterization
    missions?
\end{itemize}
Special emphasis will be put on planets in the \emph{Habitable Zone}
because this is a region of prime interest for astrobiology. Indeed
orbital parameters obtained with NEAT will allow spectroscopic
follow-up observations to be scheduled precisely when the
configuration is the most favorable.

\subsection{High-precision differential astrometry}
\label{sec:astrometry}

\begin{figure}[t]
  \centering
  \includegraphics[width=\hsize]{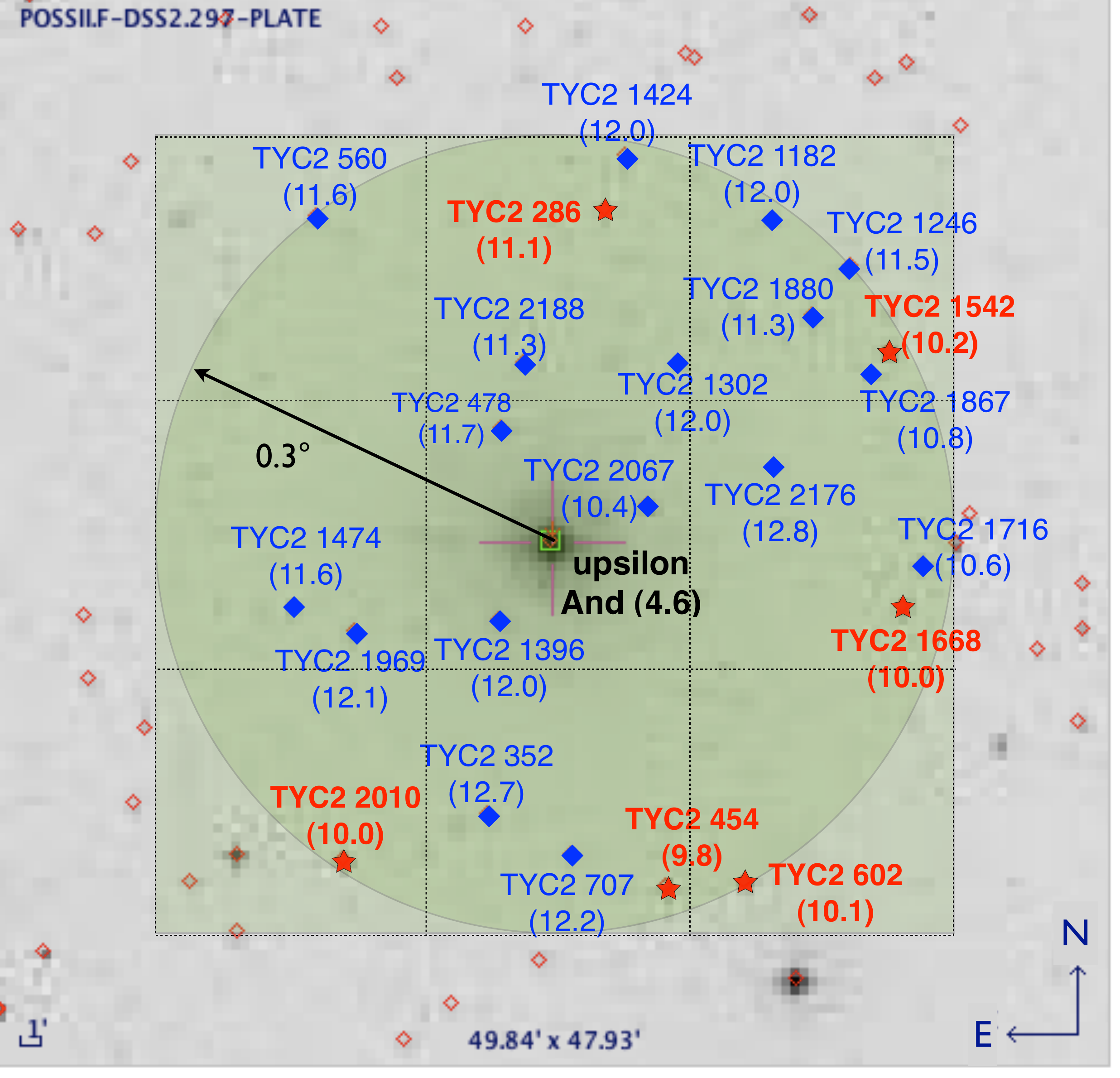}
  \caption{$0.3^\circ$ stellar field around upsilon Andromedae, a
    proposed NEAT target. There are six possible reference stars in
    this field marked in red (five $V<11$ stars
    and a $V=11.1$ one).}
  \label{fig:ups-and}
\end{figure}

\begin{figure*}[t]
  \centering
  \includegraphics[width=\hsize]{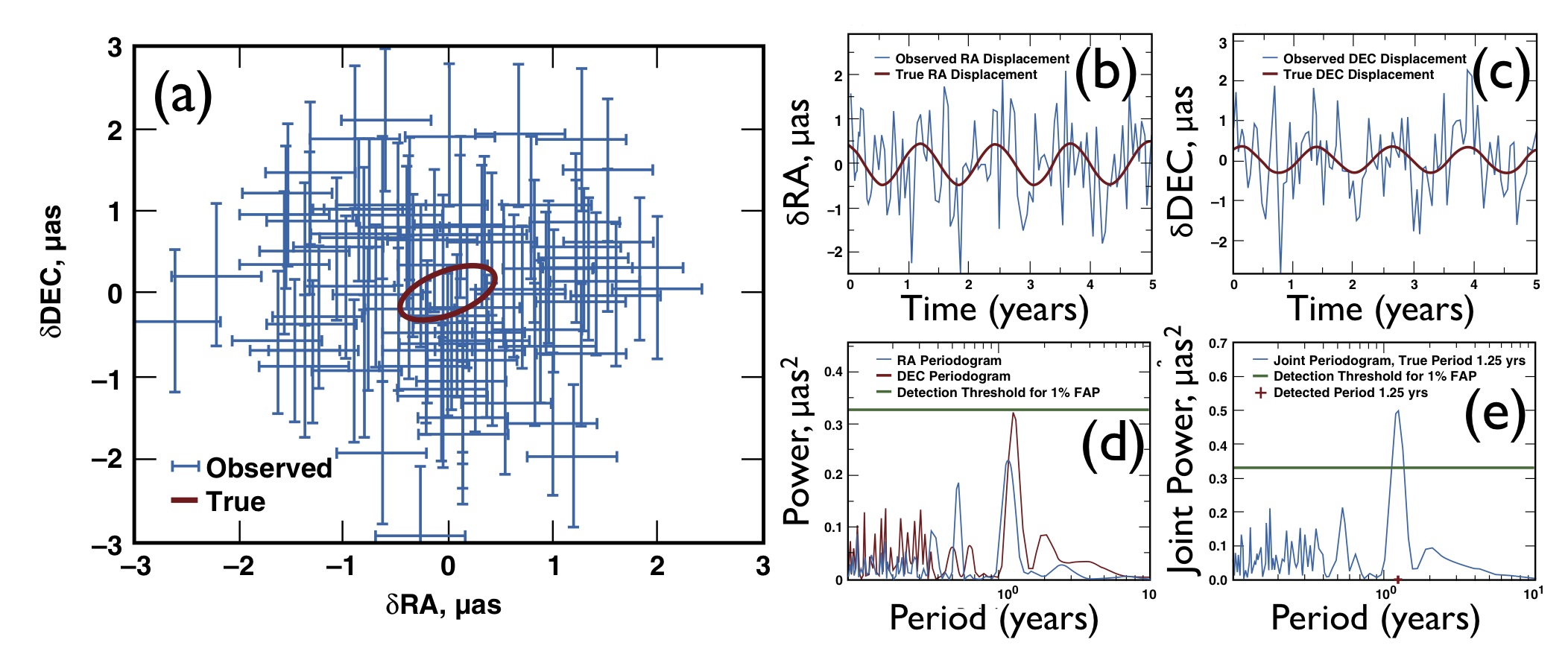}
  \caption{Simulation of astrometric detection of a planet with 50
    NEAT measurements (RA and DEC) over 5 yrs. Parameters are:
    $\Mplanet=1.5\,\MEarth$, $a=1.16\,AU$, $\Mstar=1\,\Msun$,
    $D=10\,\pc$, $\SNR\geq6$. (a) Sky plot showing the astrometric
    orbit (solid brown curve) and the NEAT measurements with error
    bars (in blue); (b) and (c) same data but shown as time series of
    the RA and DEC astrometric signal; (d) Separated periodogram of RA
    (blue line) and DEC (brown line) measurements. (e) Joint
    periodogram for right ascensions and declinations
    simultaneously. Whereas the orbit cannot be determined from the
    astrometric signal without the time information, its period is
    reliably detected in the joint periodogram (1.25\,yr) with a
    false-alarm probability below 1\% (green line). Then, the
    planetary mass and orbit parameters can be determined by fitting
    the astrometric measurements.}
    \label{fig:earth-detection}
\end{figure*}

The principle of NEAT is to measure accurately the offset angles
between a target and 6-8 distant reference stars with the aim of
differentially detecting the reflex motion of the target star due to the
presence of its planets. An example of a field that will be
observed is shown in Fig.~\ref{fig:ups-and} and a simulation of what
will be measured is displayed in Fig.~\ref{fig:earth-detection}.

The output of the analysis is a \emph{comprehensive} determination of
the mass, orbit, and ephemeris of the different planets of the
\emph{multiplanetary system} (namely the 7 parameters $\Mplanet$, $P$,
$T$, $e$, $i$, $\omega$, $\Omega$), down to a given limit depending on
the star characteristics, e.g.\ 0.5, 1 or 5\,\MEarth. The astrometric
amplitude, $A$, of a \Mstar mass star due to the reflex motion in
presence of a \Mplanet mass planet orbiting around with a semi-major
axis $a$ at a distance $D$ from the Sun is
\begin{equation}
  \label{eq:astrometric-signal}
  A = 3 \left(\frac{\Mplanet}{1\,\MEarth}\right) \left(\frac{a}{1\,\AU}\right) 
  \left(\frac{\Mstar}{1\,\Msun}\right)^{-1}
  \left(\frac{D}{1\,\pc}\right)^{-1}  \uas. 
\end{equation}
To detect such a planet, one needs to reach a precision $\sigma =
A/\SNR$ with a typical signal-to-noise ratio\footnote{Simulations
  like the ones presented in Fig.~\ref{fig:earth-detection} show that
  $\SNR=5.8$ results in a false alarm probability of 1\%.} $\SNR=6$. If
$\sigma_0$ is the precision that NEAT can reach in one single
observation that lasts $t_0$ (e.g.  $\sigma_0 = 0.8\,\uas$ in $t_0 =
1$\,h), when observing the same source \Nvisits times during \Tvisits
each visit requires
\begin{equation}
  \label{eq:Tvisits}
  \Tvisits = t_0 \left(\frac{A}{\SNR\,\sigma_0}\right)^{-2} (2\Nvisits - m)^{1/2}
\end{equation}
for a given \Nvisits, and with $m = 5+7p$ parameters where $p$ is the
number of planets in the system since there are 5 parameters
characterizing the star astrometric motion and 7 parameters for each
orbit. $\Nvisits \approx 50$ is sufficient to solve for the parameters
of 3 to 5 planets per system, for a 5-yr duration of the mission.

\subsection{Targets}
\label{sec:mission-definition}

\begin{table*}[t]
  \centering
  \caption{Partial list of possible targets, the full list
    is available on the NEAT
    website (\emph{\neattargetsurl}). 
    Stars are ranked by decreasing astrometric signal for a planet in its habitable zone (HZ). This
    signal $A$(\uas) is calculated for 0.5, 1 and 5\,\MEarth planets around the 5,
    70 and 200 first stars, respectively, assuming that the planet is
    located at the inner boundary of the HZ that secures its detection
    whenever the planet is in this zone. The corresponding integration
    time ($t_{\rm visit}$ in h) and cumulated times ($t_{\rm tot}$
    in h) are calculated for a detection with an
    equivalent $\SNR = 6$. The total time corresponds to 70\% of the
    available mission time with a 22\% margin.  
  }
  \smallskip
  \includegraphics[width=\hsize]{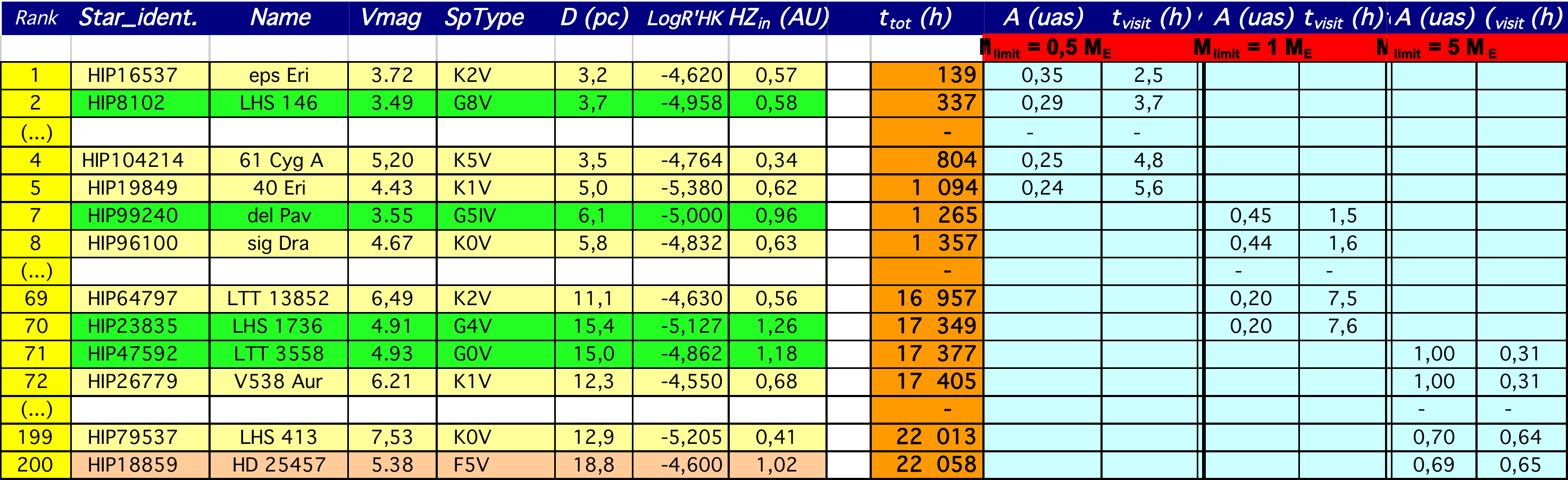}
\label{tab:target-list}
\end{table*}

\begin{figure}[t]
  \centering
  \includegraphics[width=0.6\hsize]{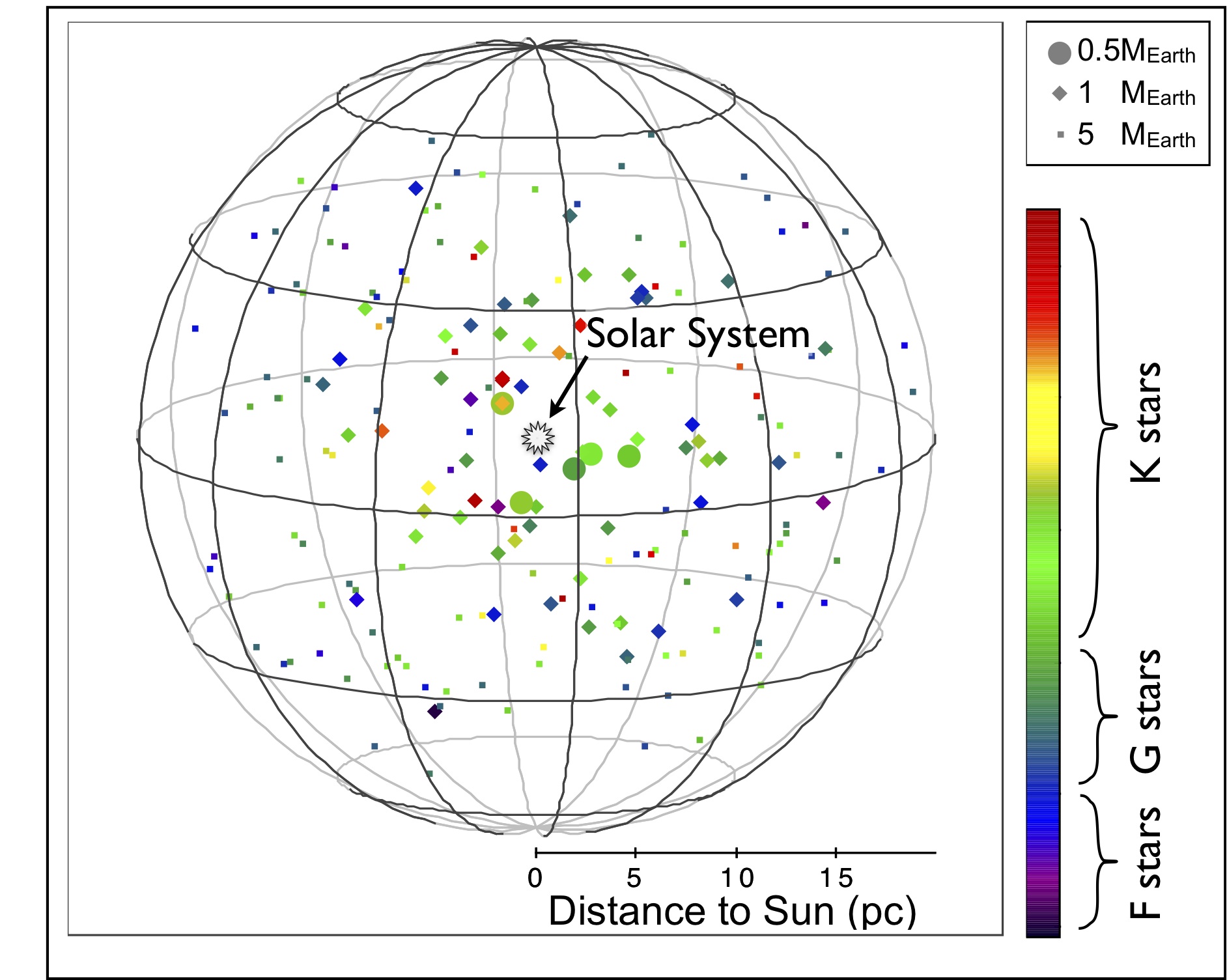}
\caption{Representation of the NEAT targets in the 3D sphere of our
    neighborhood ($D$ up to $\approx15\,\pc$). They correspond to a
    volume limited sample of all stars with spectral types between F
    and K.}
  \label{fig:targets-sphere}
\end{figure}

A possible target list is shown in Table~\ref{tab:target-list} where
we consider the list of the nearest F, G, K stars deduced from the
Hipparcos 2007 catalogue \citep[new data reduction
by][]{2007A&A...474..653V}, disregarding spectroscopic binaries, and
stars with an activity level 5 times greater than that of the Sun because of their
astrometric noise \citep[only 4\% of the
sample,][]{2011A&A...528L...9L} and for which we compute the
astrometric signal for a planet with given mass in the HZ of the stars
\citep{2010AsBio..10..103K}. Conservatively, we select the inner part
of the HZ in order to be able to detect the planet whatever is its
location in the HZ. The required number of visits and cumulative time
to observe this list of target stars is summarized in Table~\ref{tab:program}.
\begin{table}[t]
  \centering
  \caption{Left: summary of the main program capabilities and required
    resources. Right: Time and allocated maneuvers for the different
    programs: (1) the Gaia Mission and its Exoplanet Science Potential;
    (2) NEAT follow-up program of Gaia detected  planetary systems;
    (3) observations of young stars; and (4) characterizing planetary
    systems around some of the closest M stars.} 
  \label{tab:program}
  \smallskip
    \includegraphics[width=0.46\hsize]{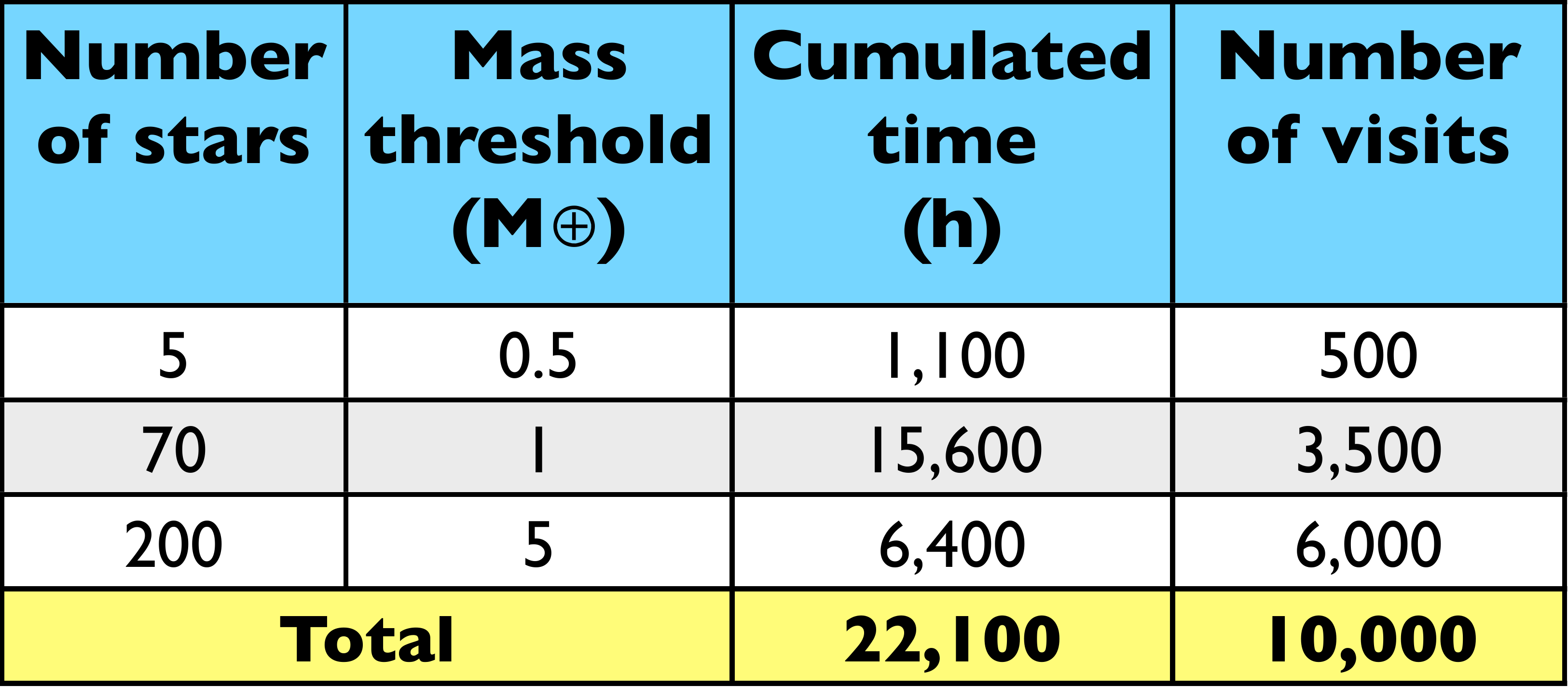}
    \includegraphics[width=0.5\hsize]{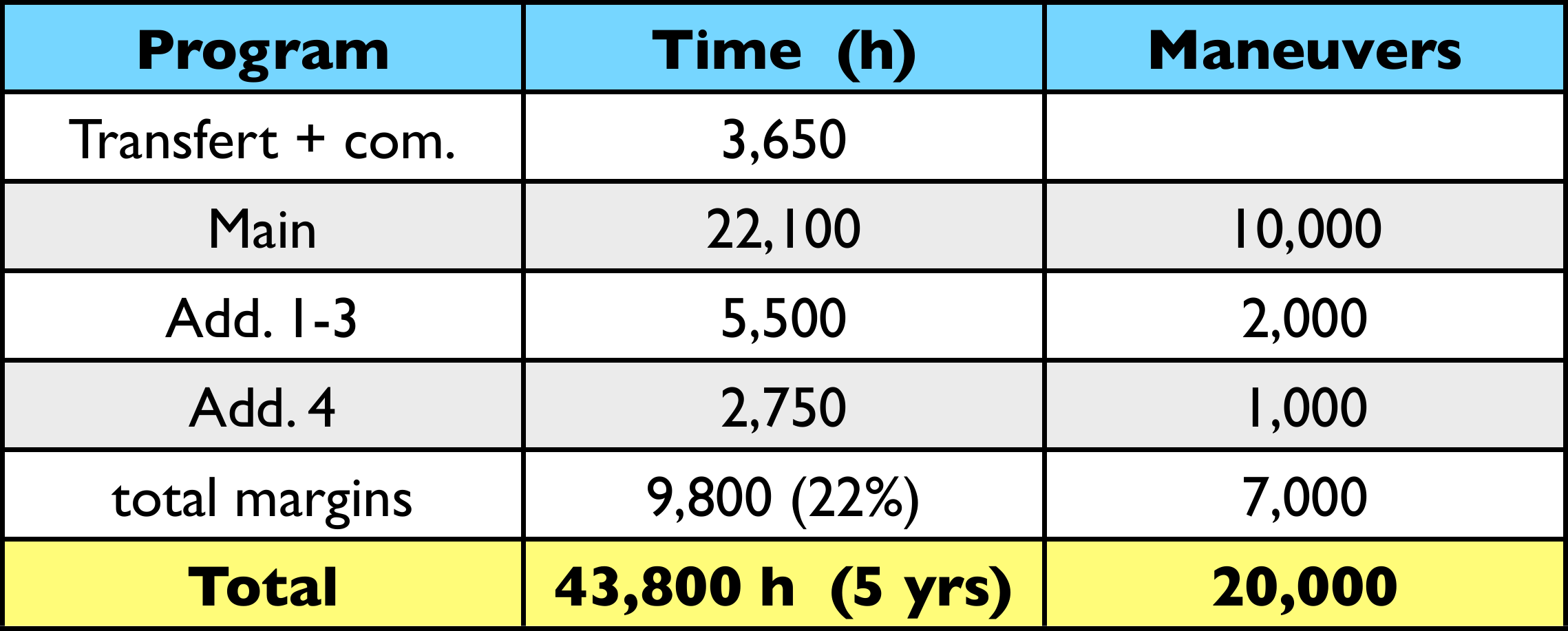}
\end{table}
The list corresponds to an exhaustive search for 1 Earth mass planets
(resp.\ 5 Earth mass planets) around K stars up to 6 pc (resp.\ 12
pc), G stars up to 10 pc (resp.\ 17 pc), and F stars up to 14 pc
(resp.\ 19 pc) in the whole HZ of the star, excluding spectroscopic
binaries and very active stars. The spatial repartition of targets is
shown in Fig.~\ref{fig:targets-sphere}. 60\% of the NEAT targets (118)
are brighter than $V=6$ and therefore will not be investigated by Gaia
because of its bright limit. So, even if some of those sources do not
harbor Earth-like planets, NEAT will be contributing to the
improvement of our knowledge about the neighborhood of our Solar
System.
In that respect, NEAT observations will not only be complementary to Gaia's
ones, but NEAT data will also form a base to improve Gaia results.

In addition to the survey for the NEAT main science program,
we propose that 30\% of NEAT time is allocated to study some
objects of interest (planets around M dwarfs, young stars, multiple
systems,... discovered by Gaia and others). The global required amount
of time and number of maneuvers is listed in Table~\ref{tab:program}
(right part).

\section{NEAT concept}
\label{sec:neat-concept}

Our goal is to detect the signal corresponding to the reflex motion of
a Sun-like star at 10\,pc due to an Earth-mass planet in its HZ, with
an equivalent final SNR of 6. That astrometric signal is
0.30\,\uas. The required noise floor is 0.05\,\uas, over 100 times
lower than Gaia's best precision (7\,\uas).  

\subsection{Technical challenges}
\label{sec:technical-challenges-1}

Achieving sub-micro-arcsecond astrometric precision, e.g. 0.8\,\uas,
in 1 hour and a noise floor under 0.05\,\uas with a telescope of
diameter D requires mastering all effects that could impact the
determination of the position of the point spread function. The
typical diffraction limited size of an unresolved star is about $1.2\lambda/D$,
which corresponds to 0.16 arcseconds for a 1-m telescope operating in
the visible spectral region. The challenge is therefore to control these
systematics effects to a level better than 1 part in 3 million
($1:3\times10^6$). 
Even though differential astrometry of stars within the same field of
view softens somewhat the requirement, this level of accuracy can only
be obtained in an atmosphere-free space environment.

Sub-micro-arcsecond level astrometry requires solutions to four
challenges:
\begin{itemize}
\item \textbf{Photon noise.} Most targets are $R\leq6$\,mag stars, but
  the required reference stars are $R\leq11$\,mag so they dominate the
  photon noise.  Using the mean stellar density in the sky, one finds
  that a field of view (FOV) as large as diam $0.6^\circ$ is needed to
  get several (6 to 8) of theses references (see e.g.\
  Fig~\ref{fig:ups-and}).
\item \textbf{Beam walk.} A classical three mirror anastigmat (TMA)
  telescope can also manage a $0.6^\circ$ diffraction limited
  FOV. However the light coming from different stars, and therefore
  from different directions, will hit the secondary and tertiary
  mirrors on different physical parts of the mirrors. The mirror
  defects will therefore produce different and prohibitive astrometric
  errors between the images of the stars. Using a single mirror
  telescope solves this problem. To obtain sufficiently high angular
  resolution, a long focal length ($\approx40$\,m) for this mirror is
  needed, with no intermediate mirrors, a relatively unusual solution
  in modern optical astronomy.
\item \textbf{Stability of the focal plane.} Proper Nyquist sampling
  with typical detector pixels of the order of 10\,\microns requires a
  focal plane at a focal length of 40\,m. Such a focal plane covering
  a FOV of $0.6^\circ$ diameter would yield a costly detector mosaic
  with $40,000\times40,000\approx10^9$ pixels. Sub-microarcesc
  astrometry over a $0.6^\circ$-diameter FOV requires the geometry of
  the focal plane to be stable to
  $\approx1:2\times10^{-10}$. Therefore thermal stability of the focal
  plane geometry will be a major challenge although it has to be
  investigated in details. Instead of building a gigapixel focal plane
  with unprecedented stability we plan to use 9 small $512\times512$
  CCDs (Fig.~\ref{fig:focal-plane-concept}) and a laser metrology
  system to measure the position of every pixel to the required
  precision, once every 10 to 30\,s. We do not rely on their
  positioning, but measure it accurately with a laser metrology based
  on dynamic interference fringes.
\item \textbf{Quantum efficiency (QE) variations.} The dynamic fringes
  also allow the measurement of the inter- and intra-pixel QE
  variations. We characterize each pixel response with six parameters
  such that the systematic errors are kept below
  $10^{-6}$. This is a process derived from the SIM studies.
\end{itemize}
\begin{figure}[t]
  \centering
  \includegraphics[width=\hsize]{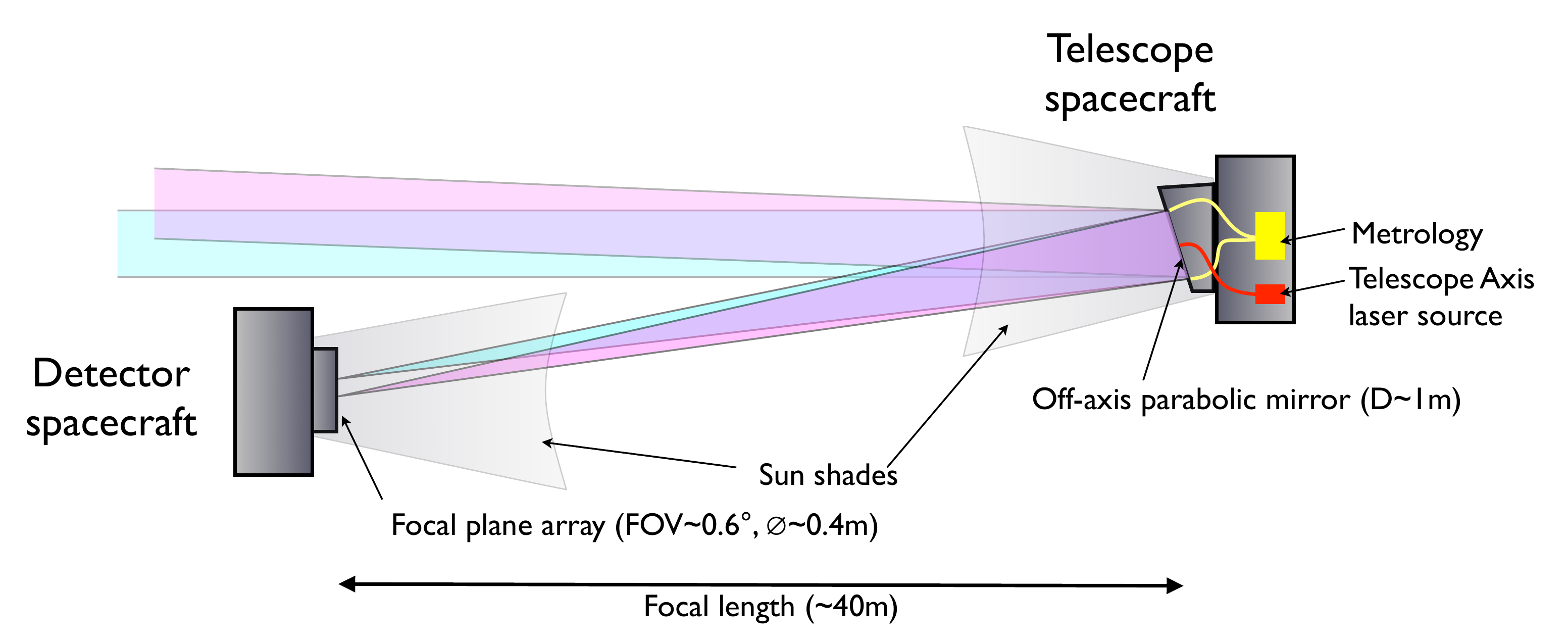}
  \caption{Proposed concept for a very high precision astrometry
    mission. It consists in two separated modules, the first one
    carrying the primary mirror (upper right) and the second one the
    detector plane (bottom left).}
  \label{fig:payload-concept}
\end{figure}
These effects have in the past hampered the performance of space missions like HST. 

\subsection{Instrumental concept}
\label{sec:instrumental-concept}

\begin{figure}[t]
  \centering
  \includegraphics[width=0.6\hsize]{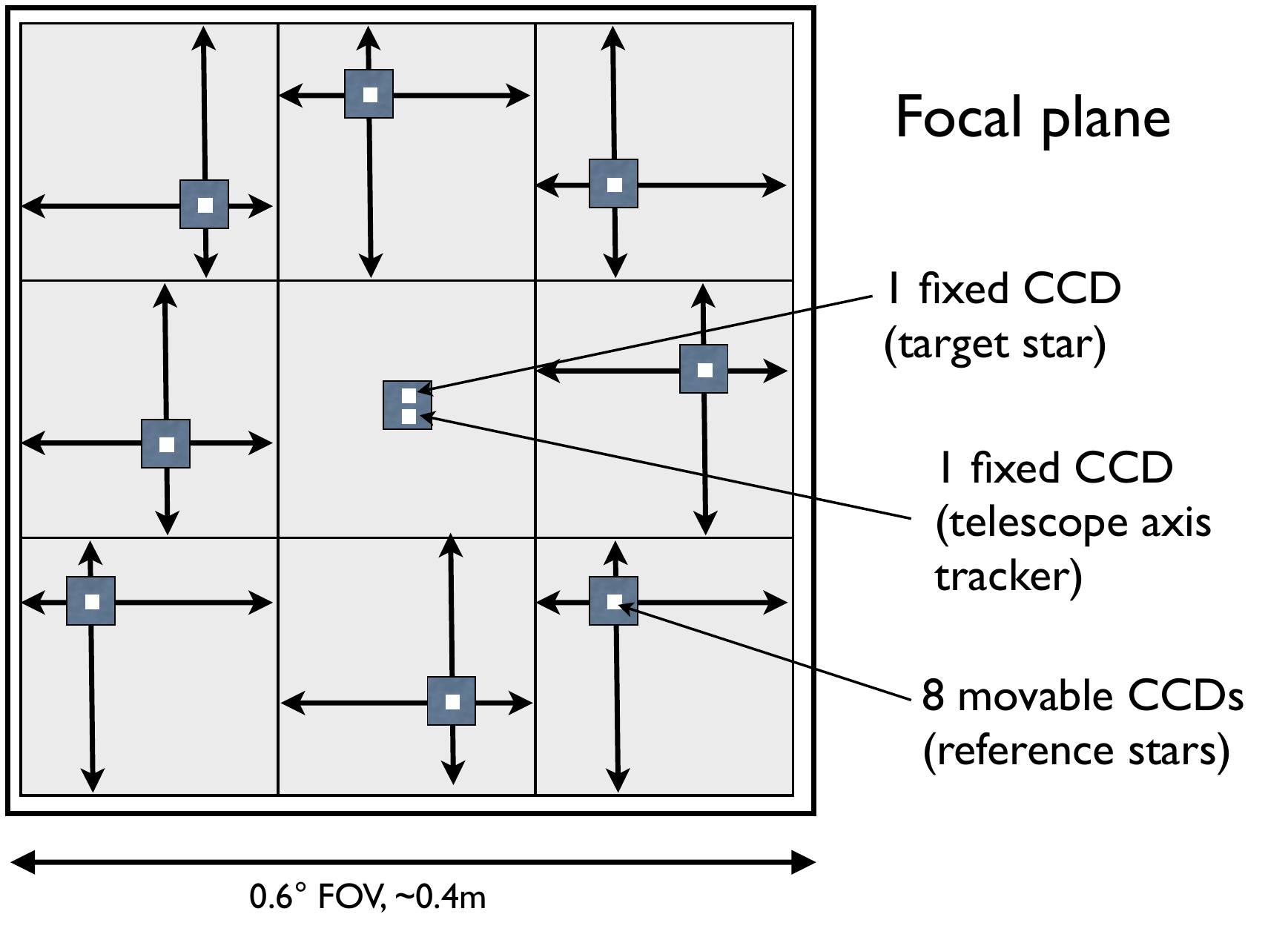}  
  \caption{Schematic layout of the focal plane. The field of view is
    divided in $3\times3$ sub-fields. Exterior subfields have visible
    arrays which can be moved in X and Y directions to image the
    reference stars. The central field has two fixed arrays, one for
    the target star and one for the telescope axis tracker.}
  \label{fig:focal-plane-concept}
\end{figure}

The proposed mission is based on a concept recently proposed by
M.~Shao and his colleagues that results from the experience gained in
working with many astrometry concepts (SIM, SIM-Lite,
corono-astrometry\footnote{See \citet{2010SPIE.7731E..68G}}).  The
concept is sketched in Fig.~\ref{fig:payload-concept} and consists of
a primary mirror ---an off-axis parabolic 1-m mirror--- a focal plane
40 m away, and metrology calibration sources. The large distance
between the primary optical surface and the focal plane can be
implemented as two spacecraft flying in formation, or a long deployed
boom. The focal plane with the detectors having a field of view of
$0.6^\circ$ is shown in Fig.~\ref{fig:focal-plane-concept}. It has a
geometrical extent of $0.4\meters\times0.4\meters$. The focal plane is
composed of eight $512\times512$ visible CCDs located each one on an
XY translation stage while the central two CCDs are fixed in position.
The CCD pixels are 10\microns in size.

\begin{figure*}[t]
  \centering
  \includegraphics[trim= 0em 8em 6em 6em, clip,width=\hsize]{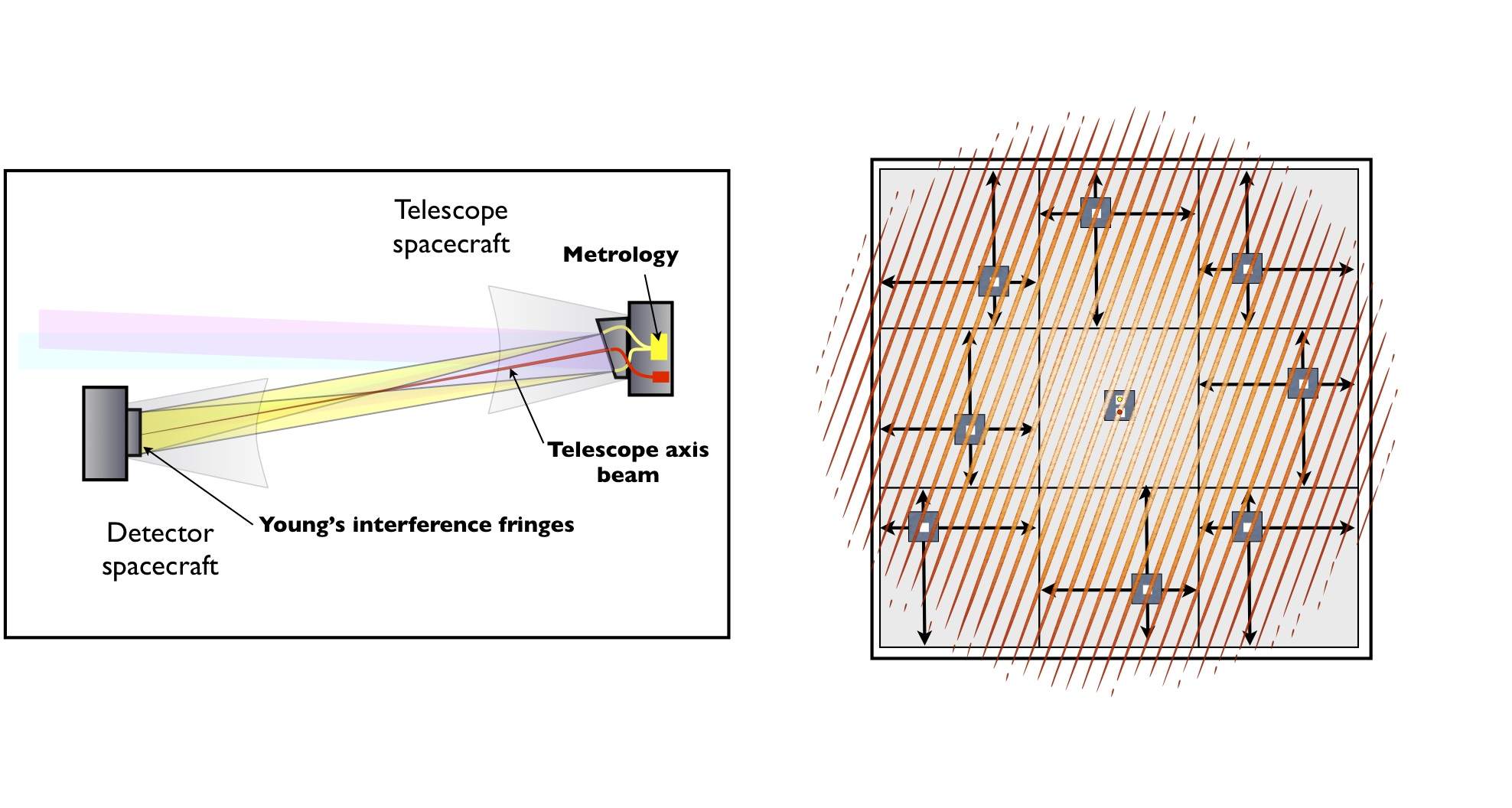}
  \caption{Principle of the metrology and the axis tracker. Left
    panel: the metrology laser light (in yellow) is launched from
    fibers located at the edge of the mirrors. Right panel: the laser
    beams interfere over the detector plane. Only the fringes
    corresponding to a pair of fibers are represented on this figure
    and they are not to scale, since the fringe spacing is equal to
    the PSF width. The axis tracker (sketched in red on the left
    panel) is a laser beam launched in the center of the mirror that
    is monitored in the lower central CCD.}
  \label{fig:metrology-concept}
\end{figure*}

The principle of the measurement is to point the spacecraft so that
the target star, which is usually brighter ($R\leq6$) than the
reference stars ($R\leq11$), is located on the axis of the telescope
and at the center of the central CCD. Then the 8 other CCDs are moved
to center each of the reference stars on one of them. To measure the
distance between the stars, we use a metrology calibration system that
is launched from the telescope spacecraft and that feeds several
optical fibers (4 or more) located at the edge of the mirror. The
fibers illuminate the focal plane and form Young's fringes detected
simultaneously by each CCD (Fig.~\ref{fig:metrology-concept}). The
fringes have their optical wavelengths modulated by acoustic optical
modulators (AOMs) that are accurately shifted by 10\,Hz, from one
fiber to the other so that fringes move over the CCDs.  These fringes
allow us to solve for the XYZ position of each CCD. An additional
benefit from the dynamic fringes on the CCDs is to measure the QE of
the pixels (inter-, and intra-pixel dependence). The CCDs are read at
50\,Hz providing many frames that will yield high accuracy.

With the proposed concept, it is possible to achieve all of the main technical
requirements:
\begin{itemize}
\item \textbf{Focal plane stability.} Instead of maintaining a focal
  plane geometry stable at the 0.1\,nm level for a 5-yr duration,
  which is impossible, we implement a metrology for every pixel at the
  sub-nanometer level, with an interferometric system that has been
  qualified by the SIM-Lite laboratory demonstrators.
\item \textbf{Reference frame.}  By measuring the fringes at the
  sub-nanometer level using the information from all the pixels of
  each CCD (SIM-Lite technology), it is possible to solve for the
  position of all reference stars compared to the central target with
  an accuracy of 0.8\,\uas per hour. The
  field of view of $0.6^{\circ}$ allows us to have 6 to 8 reference stars
  brighter than $V=11$ in most fields.
\item \textbf{Photon noise.} The field of $0.6^{\circ}$ provides about 6 to
  8 stars of magnitude brighter than $R=11$. The number of photons
  received by one 11-mag star on the system is $\approx
  4.1\times10^9\,\mbox{ph/hr}$. Since the FWHM of diffraction-limited
  stars is $1.2\lambda/D=0.16$ arcsecond, the photon noise limit in
  1\,h of integration due to a set of 6 reference stars is
  $(\lambda/2D)/\sqrt{6N}\approx 0.5\,\uas$.  With more than
  $50\times2$ measurements of a few hours spread over 5\,yrs, the
  equivalent precision is 0.05\,\uas in RA and Dec, corresponding to
  the detection of the 0.30\,\uas signal with a SNR $\approx 6$.
\item \textbf{Large-scale calibration.} The detector plane does not
  have to be fully covered by pixels, since the positions of the
  reference stars are known from available catalogues (10-20\,mas for
  Tycho 2, and about tens of \uas for Gaia). For the target stars
  ($R\leq6$), we use the Hipparcos catalog (few mas accuracy). This
  corresponds to $<1/10^{\rm th}$ of the PSF or the fringe width. The
  number of fringes between the target star and the reference stars is
  then known, only the positions of the star centroids relative to the
  interferometric fringe have to be measured accurately.
\end{itemize}
The use of 10 small CCDs drastically reduces the cost of what would
otherwise be a giga-pixel focal plane and also helps to control
systematics. With such a concept, the mission performance
would be similar to, and even more favorable for exoplanets, than what
was proposed for SIM-Lite with 5 years of operation, but at the price
of giving up all-sky astrometry and the corresponding science
objectives.

\subsection{Performance assessment and error budget}
\label{sec:perf-assessm-error}

Achieving a relative precision of $2\times10^{-10}$ is slightly better
than the precision achievable by only the combination of our metrology
laser, the thermal expansion coefficient of the primary mirror and our
expected temperature stability. Achieving our target precision relies
on not only the metrology stability, but also on the precise knowledge
of the positions of the multiple reference stars used since the
expected motions of the references cannot be considered as fixed (see
discussion in Sect.~\ref{sec:astro-challenges}). Our comprehensive error budget takes into
account all sources of error, including instrumental effects, photon
noise and astrophysical errors in the reference star positions.

\begin{figure*}[t]
  \centering
  \includegraphics[width=\hsize]{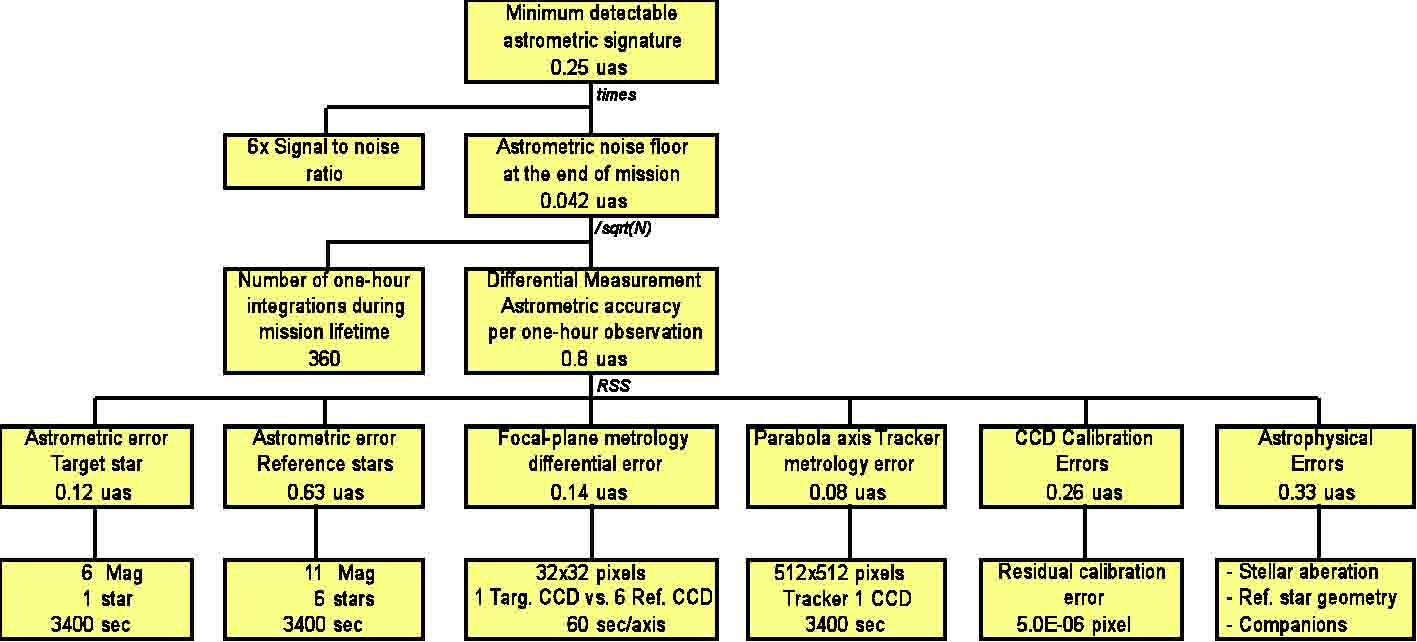}
 \caption{Top-level error budget for NEAT. It shows how the 0.8\,\uas
    accuracy enables the detection of 0.3\,\uas signatures with a signal
    to noise of 6 after 360\,h of observation. It also shows the
    major contributors to the astrometric error.} 
  \label{fig:error-budget}
\end{figure*}

The biggest term is the brightness dependent error for the set of
Reference stars. The half-width of the PSF for the coma-aberrated
images of the reference stars is about 19\,\microns on the focal plane
(or 100\,mas on the sky). After 1\,s of integration, $1.3\times10^6$
photoelectrons are detected for each of the 11-mag reference star;
their centroid location can be estimated to $0.016\,\microns$ rms
($1.6\times10^{-3}$ pixel or 0.08\,mas). Since all the stars are
measured simultaneously, the stars do not need to be kept centered on
the detector at the sub-mas level, but only to a fraction of the PSF
width to avoid spreading of the photon outside of the PSF and
therefore cause the PSF effective width to be larger. A tenth of pixel
(1\,\microns) stability over the one-second integration is sufficient.
After 3400\,s of integration, the statistical averaged position of the
barycenter of the set of reference stars ($R\leq11$\,mag) will be
measured with a residual 0.126\,nm (0.63\,\uas) uncertainty.
Similarly, the position of the target star ($R\leq16$\,mag) will be
measured with a residual 0.024\,nm (0.12\,\uas) uncertainty. Although
the spacecraft will have moved by several arc-seconds, the
differential position between the target star and the barycenter of
the set of reference stars will be determined to 0.64\,\uas.

Similarly, the focal plane metrology system will have determined the
differential motion of the target CCD relative to the barycenter of
the set of reference CCDs with an error smaller than 0.16\,\uas after
$60\times 1$\,s metrology measurements.

NEAT will not be capable of measuring the absolute separation between
the target and the set of reference stars to 0.8\,\uas. NEAT
objectives will therefore be to measure the change in the \emph{relative}
position of those stars between successive observations spread over
the mission life, with an error of 0.8\,\uas for each one hour visit.
The six major errors terms are captured in the simplified version of the
error budget shown in Fig.~\ref{fig:error-budget}.

If unmonitored, the displacement of the projected field aberrations on
the focal plane would produce a 60\,\uas differential astrometric error
per arcsecond of relative spacecraft motion. The telescope axis
tracker will monitor the relative position of the focal plane relative
to the parabola axis simultaneously with the stellar observation with
a 1\,mas accuracy per hour. This will be sufficient to correct the
observations during post-processing for the field-dependent aberration
to better than 0.1\,\uas.

Static figure errors of the primary mirror will produce centroid
offsets that are mostly common-mode across the entire field of
view. Differential centroid offsets are significantly smaller than the
field-dependent coma and are in fact negligible.  Similarly, changes
in the primary mirror surface error, e.g.\ due to thermal
dilatation\footnote{The coefficient of thermal expansion of the mirror
is about 100 times smaller than those of the elements that compose
the detector. The metrology parameters are constantly monitored.},
meteorite impacts,... produce mostly common-mode centroid shifts and
negligible differential centroid offsets. On the other hand,
displacement and changes in the shape of the PSF would couple with the CCD
response if the CCD response is not properly calibrated. This is
continuously done by the metrology fringes.

\subsection{Design of the payload subsystems}
\label{sec:design-payload-subsystems}

\begin{figure*}[t]
  \centering
  \includegraphics[trim=20mm 20mm 20mm 0mm,clip, width=\hsize]{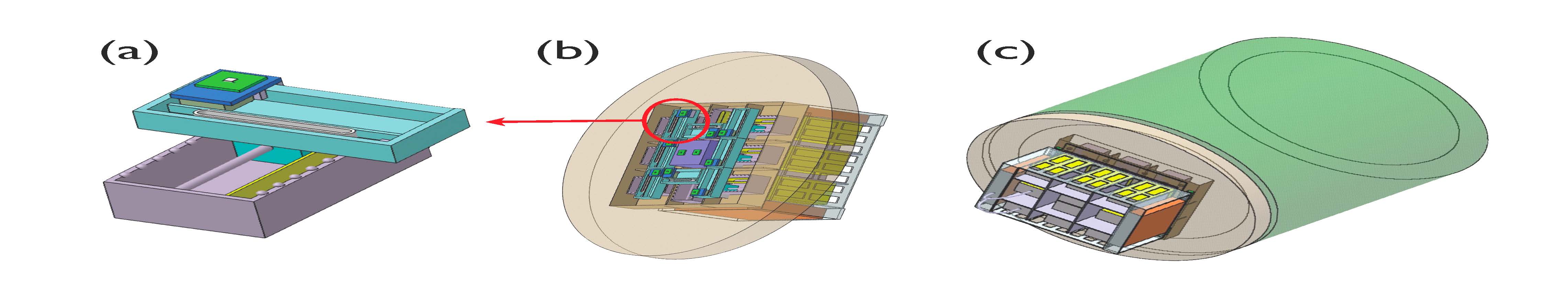} 
 \caption{Views of the focal plane assembly. (a) Magnified
    view of one of the 8 XY translation stages of the focal plane. In
    yellow, the $512\times512$ CCD and its support. In blue and green the two
    translation stages. (b) The front part of the focal plane
    with its 8 movable CCDs and two fixed CCDs at the center. (c) The electronics racks.}
  \label{fig:focal-plane-drawing}
\end{figure*}

\textbf{Focal plane assembly}. A proposition for implementation of the focal plane is shown in
Fig.~\ref{fig:focal-plane-drawing}.
The detector is foreseen to be a CCD fabricated by the E2V
technologies company in UK. The target star, the reference star and
the telescope axis tracker will all use the same CCD that includes
the capability to read windowed images, typically $10\times10$ to
$30\times30$ pixels. The 8 XY tables consist of two linear tables
mounted on top of each other. Each table uses a piezo-reptation
motor\footnote{Such reptile motors have been qualified by the Swiss
  firm RUAG for the LISA GPRM experiment.}, a linear ball bearing
system and an optical incremental encoder. These motors fulfill
several requirements of simplicity: they are self-locked when they are
not powered; they can be used both for large displacements by stepping up to
$100\,\mbox{mm}\times100\,\mbox{mm}$ and elementary analog motion down
to 50\,nm.  Since 8 tables are used in parallel in the focal plane,
the loss of one table is not a single point failure. An
alternative implementation could be to drive the XY tables with ball
screws and rotary motors. The limited resolution of such a motor stage
(about 5\,\microns) could be supplemented by a second high-resolution
piezo XY table\footnote{such as the Cedrat XY25XS}, mounted on top of the first
XY table.  The main structure of the focal XY tables consists of a
large lightweight aluminum cylinder which is thermally controlled and in which
pockets are machined for the fixation of the XY tables.

\textbf{Telescope}. The primary mirror is an off-axis paraboloid, with a 1\,m diameter
clear aperture, an off-axis distance of 1\,m and a focal length of
40\,m. It would be fabricated in either Zerodur or ULE, 70\%
light-weighted and weight about 60\,kg. The surface quality should be
better than $\lambda/4$ peak-to-valley and would be coated with
protected aluminum. A 50\,mm hole at the center of the mirror
accommodates the beam launcher for the telescope axis tracker.  The
three bipods on the back of the mirror support the mirror with minimum
deflection. The bipods interface with the tip-tilt stage made of 3
preloaded piezo stacks on parallel flexures that provide the +/-6
arcsecond amplitude for two-axis articulations. The entire primary
mirror assembly interface to the telescope payload plate is a 34\,kg
hogged-out aluminum plate. This plate also hosts the metrology source,
the telescope drive electronics, the telescope baffle and the
interface to the spacecraft.
\begin{figure}[t]
  \centering
    \includegraphics[width=\hsize]{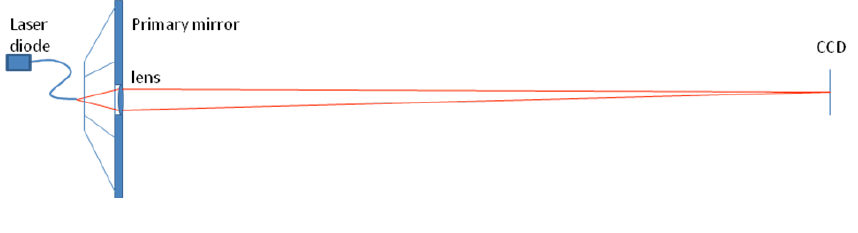} 
 \caption{Laser metrology of the telescope axis tracker. The launcher
    embedded in the primary mirror substrate forms a 600\,\microns
    (60 pixels) FWHM diffraction limited image of the fiber output on
    the Telescope Axis Tracker CCD. By sampling the entire $512\times512$
    pixel detector, the parabola axis can be tracked over a 26 arcsec range,
    with a 0.3\,mas estimation error at 10\,Hz.} 
  \label{fig:telescope-axis-tracker}
\end{figure}
The telescope axis tracker is used to estimate the location of the
primary mirror axis with respect to the focal plane in order to
monitor it and then correct for the telescope field dependent errors. The
sensor is the second fixed CCD located in the focal-plane. The
launcher would consist of either an achromatic doublet or an
aspherical singlet lens, embedded in the primary mirror substrate at
its center and a single-mode fiber-coupled laser diode. Fig.~\ref{fig:telescope-axis-tracker}
shows how the fiber tip is re-imaged onto telescope axis tracker CCD in
the focal plane.

\textbf{Laser metrology}. The focal-plane metrology system consists of
the metro\-logy source similar to the one developed for SIM
\citep{2010SPIE.7734E..71E}, the metrology fiber launchers and the
focal plane detectors (CCDs) which alternatively measure the stellar
signal (57\,s observations) and the metrology (1\,s per axis every
minutes). The metrology fiber launchers consist of nominally four
optical fibers attached to the primary mirror substrate. Three of them
are located around the edge of the mirror, and are used by pairs in
order to conduct three redundant measurements of the relative location
of the CCDs. The fourth fiber is located inside the clear aperture,
and is used in combination with each of the three other fibers to
produce three additional measurements during focal plane calibration
and calibrate the distance mirror-focal plane.

\textbf{Pointing servo systems}. The pointing of the telescope from
one target to the next one is accomplished by the two spacecraft in
formation flying. The target stars will be typically separated by
$10^{\circ}$. Re-pointing of the telescope will require rotation of the two
spacecraft by several degrees using reaction wheels and translation of
the telescope spacecraft by several meters using hydrazine
propulsion. Fine positioning of the focal plane relative to the mirror
is done by cold gas propulsion system, and at the end of the maneuver,
the telescope spacecraft will be oriented to better than 3 arcseconds
from the target star line of sight using star trackers and the focal
plane spacecraft will be positioned to better than 2\,mm from the
primary mirror focus. At that point, the spacecraft will maintain their
relative position to better $\pm$2\,mm in shear and in separation for
the duration of the observation. The separation does not require a
servo-loop of the payload, because its effect is only a degradation
in performance (when FWHM increases, final precision decreases in same
proportion) and is managed in the error budget.

%
During the observation, the instrument uses a tip-tilt stage behind
the primary mirror to center the target star on the $32\times32$ pixel
sub-window on the target star CCD.
Once in the $32\times32$ pixel sub-frame mode, the target star CCD is
read at 500\,Hz, and feedback control between the CCD and the tip-tilt
stage can be used to keep the star centered on the detector to better
than 5 milli-arc-second RMS (0.1 pixel RMS) for the duration of the
observation. This is the only active feedback loop in the instrument
system working at 50\,Hz; the other degrees of freedom (focal plane
tip, tilt, clocking and focal-plane-to-mirror separation) are
monitored but not corrected for in real-time.  Prior to acquisition,
the reference star CCDs will be pre-positioned to the expected
location of the reference stars using the translation stages. The XY
translation stage fine motion of the reference star CCDs at a
0.2\,\microns precision enables centering of the reference stars on
the detectors to better than a tenth of a pixel. Once the reference
stars are acquired, the translations stages are locked for the
duration of the observation.

\subsection{Mission requirements}
\label{sec:spacecrafts}

The objectives of the NEAT mission require to perform acquisitions
over a large number of targets during the mission timeline, associated
to a 40\,m focal length telescope satellite. The preliminary assessment of the
NEAT mission requirements allows to identify the following main
spacecraft design drivers. 

\textbf{Launch configuration and mission orbit.} The L2 orbit is the
preferred orbit, as it allows best formation flying performance and is
particularly smooth in terms of environment. The Soyuz launch,
proposed as a reference for medium class missions, offers satisfying
performance both in terms of mass and volume.

\textbf{Formation Flying and 40\,m focal length.}  The mission relies
on a 40\,m focal length telescope, for which the preferred solution is
to use two satellites in formation flying. The performance to be
provided by the two satellites in order to initialize the payload
metrology systems are of the order of magnitude of $\pm2$\,mm in
relative motion, and of 3 arcseconds in relative pointing, which are
typically compatible with Formation Flying Units and gyroless
AOCS\footnote{Attitude and Orbit Control System} architecture. In
addition, at L2, the solar pressure is the main disturbance for
formation flying control. As a result, surface-to-mass ratio (S/M
ratio) is the main satellite drift contributor and should be as close
as possible for the two satellites. Although satellite design can cope
with these requirements, the S/M ratio of the satellites will evolve
during the mission (because of fuel losses and sun angle). However,
the preliminary mission assessment tends to demonstrate that the S/M
difference between the two satellites can be reduced down to 20-30\%,
which is deemed compatible with mission formation flying requirements.


\textbf{Number of acquisitions and Mission $\Delta V$.}  The mission
aims at a complete survey of a large number of targets and the
maximization of the number of acquisitions will be a main objective of
the next mission phases. The mission objectives require a threshold of
20,000 acquisitions (see Sect.~\ref{sec:spacecraft-design} for
details).  In addition, the time allocation for these reconfiguration
maneuvers is quite limited, in order to free more than 85\% of mission
duration for observations.  As a result, the mission is characterized
by a large $\Delta V$ (550 to 880\,m/s) dedicated to reconfigurations,
plus allocations for fine relative motion initialization and control
using the $\mu$-propulsion system. This large number of reconfigurations
is also driving the number of thruster firing, which are
qualified to typical numbers of up to 5,000 to 50,000 with cycling as
required for NEAT.  

\textbf{Baffles and Parasitic Light.}  The mission
performance relies on the ability of the focal plane to receive only
star flux reflected by the telescope satellite. A first requirement is
to implement baffles on the two satellites, coupled by a diaphragm on
the focal plane.  In addition, all parasitic light coming from
telescope satellite reflections should be avoided, thus requiring all
bus elements to be shielded by a black cover.  Following this
preliminary satellite requirement analysis, a first simple and robust
mission concept has been identified.

\subsection{Preliminary Spacecraft Design}
\label{sec:spacecraft-design}

\begin{figure*}[t]
  \centering
  \includegraphics[width=\textwidth]{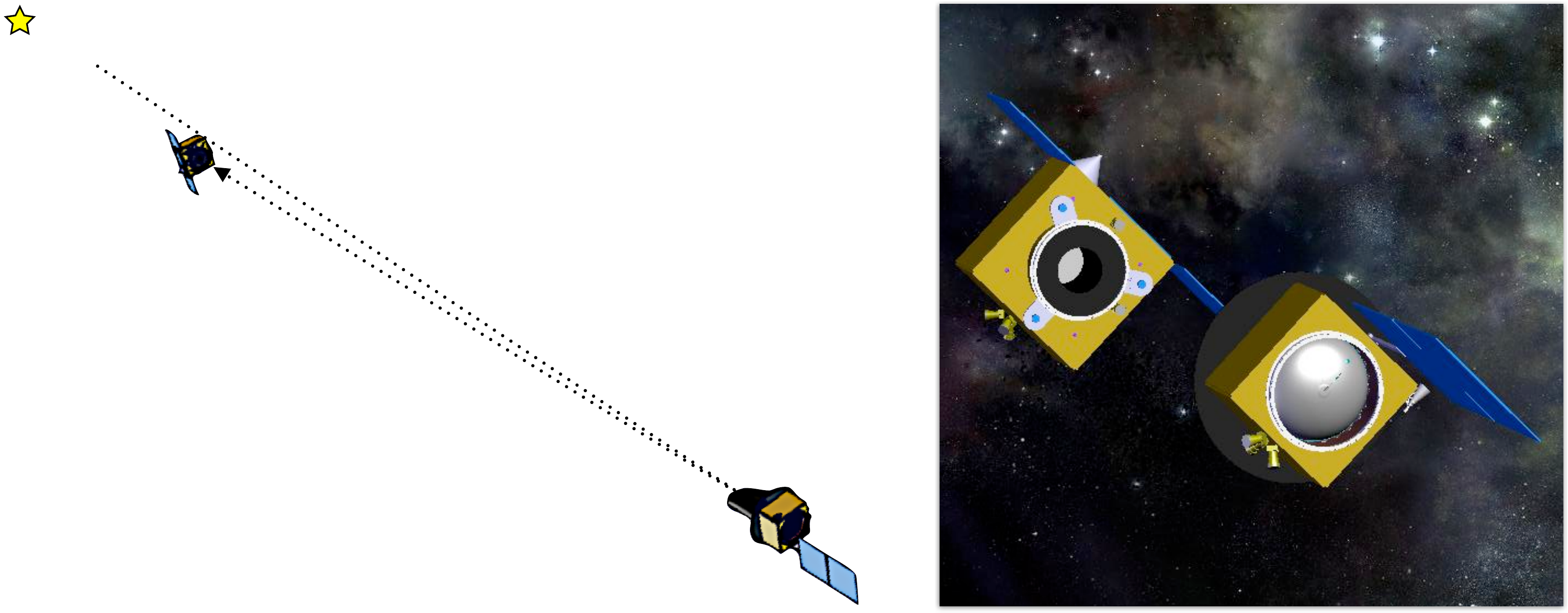}
  \caption{Left: NEAT spacecraft in operation with the two satellites
    separated by 40\,m. Right: closer external view of the two satellites.}
  \label{fig:spacecraft}
\end{figure*}

The preliminary NEAT mission assessment allowed to identify a safe and
robust mission architecture (Fig.~\ref{fig:spacecraft}), relying on
high technology-readiness-level (TRL) technologies, and leaving safe
margins and mission growth potential that demonstrates the mission
feasibility within the medium class mission cost cap.

\textbf{System Functional Description.}  The proposed mission
architecture relies on the use of two satellites in formation flying
(FF). The two satellites are launched in a stacked configuration using
a Soyuz ST launcher, and are deployed after launch in order to
individually cruise to their operational Lissajous orbit.  Acquisition
sequences will alternate with reconfigurations, during which the
Telescope Satellite will use its large hydrazine propulsion system to
move around the Focal Plane Satellite and to point at any specified
star. At the approach of the correct configuration, the Focal Plane
Satellite will use a cold gas $\mu$-propulsion system for fine
relative motion acquisition.  The Focal Plane Satellite will be
considered as the chief satellite regarding command and control,
communications and payload handling. Communications with the L2 ground
station would typically happen on a daily basis through the Focal
Plane Satellite, with data relay for TC/TM\footnote{Telecommand /
  Telemetry} from the Telescope Satellite using the
FFRF\footnote{Formation Flying Radio Frequency} units. This satellite
will however be equipped with a similar communication subsystem, in
order to support cruise and orbit acquisition, and to provide a
secondary backup link.

\textbf{Formation Flying Architecture.}  The formation flying will
have to ensure anti-collision and safeguarding of the flight
configuration, based on the successful PRISMA flight heritage. In
addition, the spacecraft will typically perform 12 to 20 daily
reconfigurations of less than $10^{\circ}$ of the system line of sight
corresponding to 7\,m of translation of one satellite compared to the
other perpendicular to the line of sight.  During these
configurations, the telescope satellite will perform translations
---supported by the FF-RF Units--- using its large hydrazine tanks
(250\,kg) for a $\Delta V \approx 605$\,m/s. When the two satellites
will approach the required configuration, the telescope satellite will
freeze, and the focal plane satellite will perform fine relative
pointing control using micro-propulsion system. As a result, the
micro-propulsion will have to compensate for hydrazine control
inaccuracies, which will require large nitrogen gas tanks (92\,kg for
$\Delta V \approx 75$\,m/s). Finally, 28\,kg of hydrazine carried by
the FP satellite allows $\Delta V \approx 55$\,m/s for station keeping
and other operations.

\textbf{Satellite Design Description.} The design of the two
satellites is based on a 1194\,mm central tube architecture, which
will allow a low structural index for the stacked configuration and
provides accommodation for payloads and large hydrazine tanks.  Strong
heritage does exist on the two satellites avionics and AOCS. In
addition, they both require similar function which would allow to
introduce synergies between the two satellites for design,
procurement, assembly, integration and tests. The proposed AOCS
configuration is a gyroless architecture relying on reaction wheels
and high-performance star trackers (Hydra Sodern), which is compatible
with a 3\,arcsec pointing accuracy (see end of
Sect.~\ref{sec:design-payload-subsystems} for payload control). The
satellites communication subsystems use X-Band active pointing
antenna, supported by large gain antenna for low Earth orbit
positioning and cruise, coupled with a 50\,W RF Transmitter. The
active pointing medium gain antenna allows simultaneous data
acquisition and downlink.  A reference solution for the satellite
on-board computer could rely on the Herschel-Planck avionics.
\begin{figure}[t]
  \centering
  \includegraphics[width=0.30\hsize]{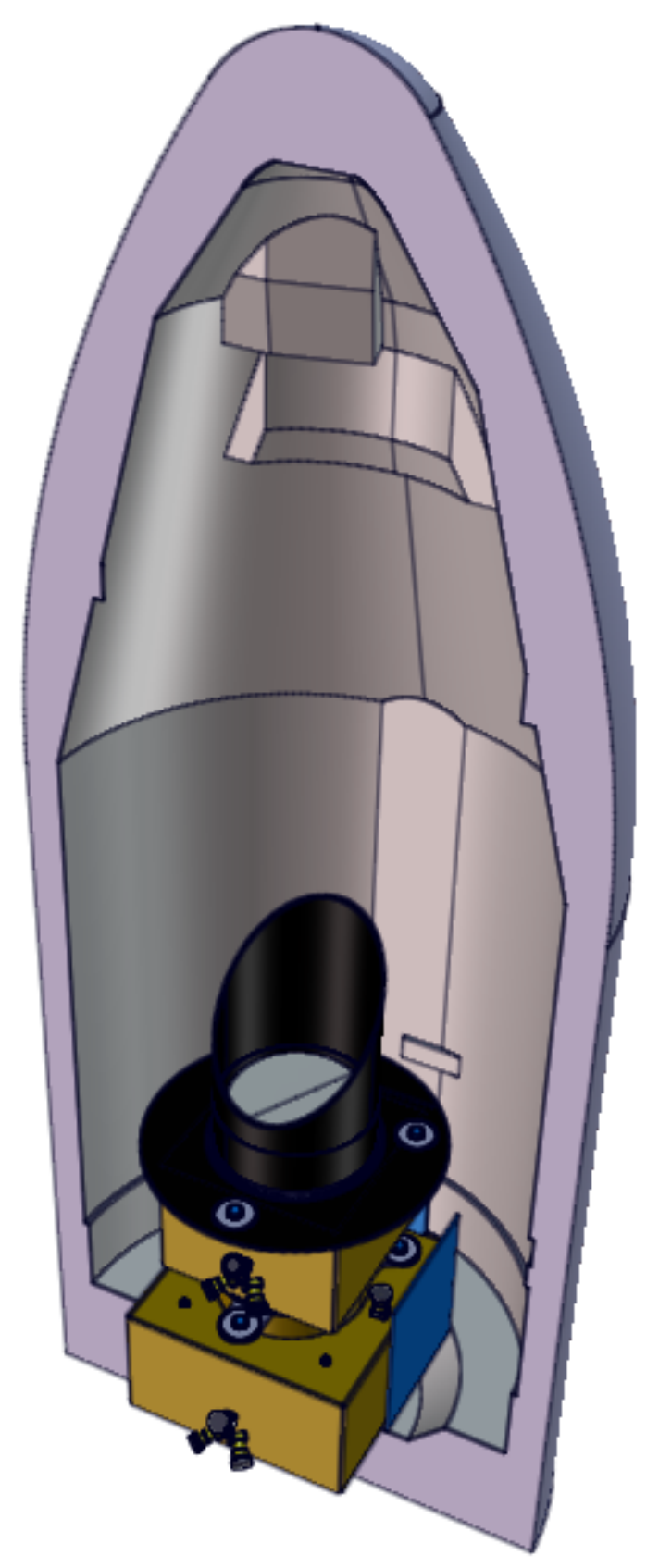}
  \includegraphics[width=0.28\hsize]{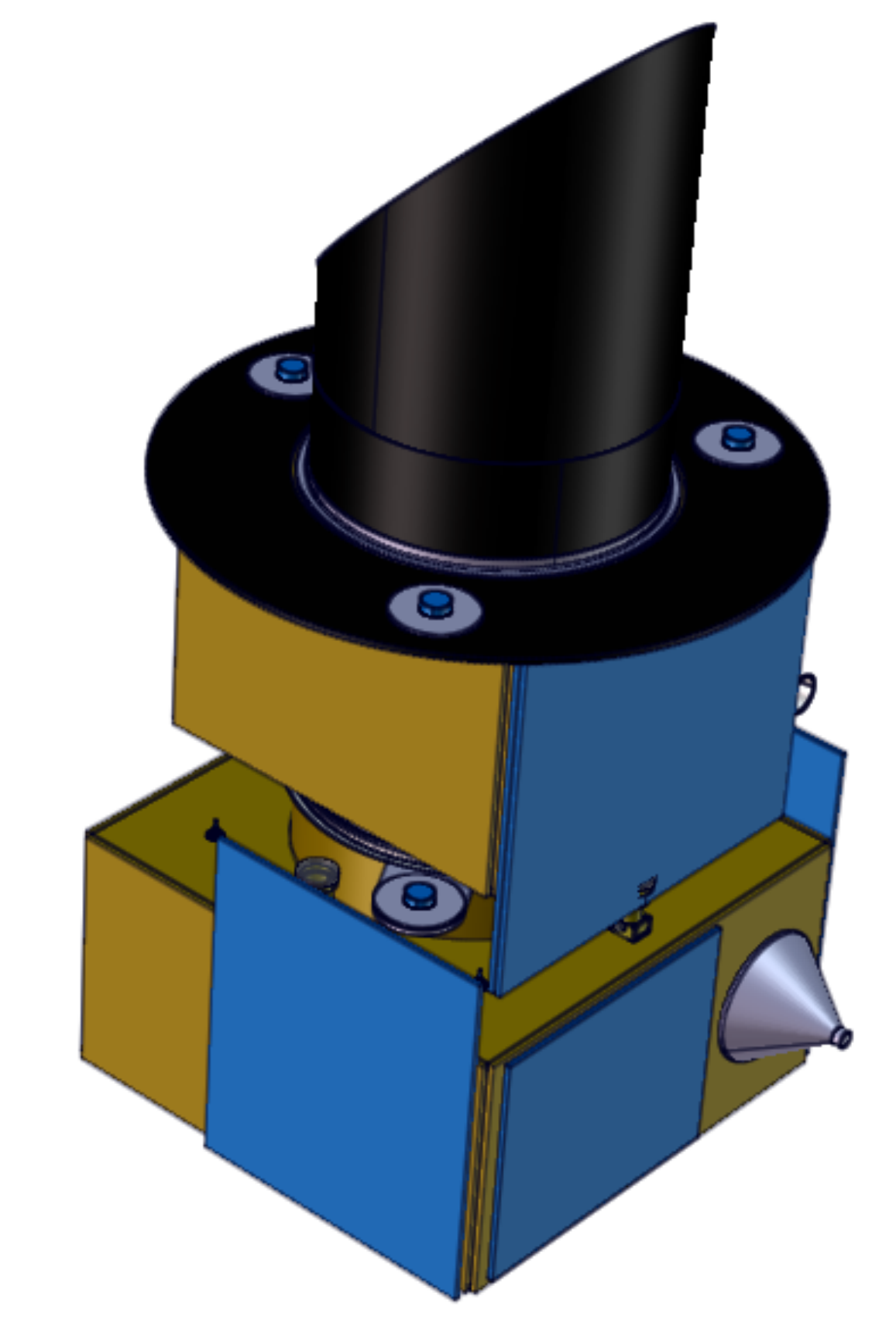}
  \caption{NEAT stowed configuration}
  \label{fig:stowed}
\end{figure}

The two satellites would have custom mechanical-thermal-propulsion
architectures. The telescope satellite features a dry mass of 724\,kg
and the focal plane satellite a dry mass of 656\,kg. The focal plane
satellite carries the stacked configuration. The payload (focal plane
+ baffle) are assembled inside a 1194\,mm central tube, which
will also ensure the stacked configuration structural stiffness. The
spacecraft bus, and large cold gas tanks, will be assembled on a
structural box carried by the central tube. The proposed architecture
uses a large hydrazine tank inside the 1194\,mm central tube which
offers a capacity of up to 600\,kg hydrazine, thus allowing both a low
filling ratio and a large mission growth potential. The payload module
---with the payload mirror, rotating mechanisms and baffle--- is then
assembled on the central tube.

\textbf{Proposed Procurement Approach.}  The NEAT mission is
particularly adapted to offer a modular spacecraft approach, with simple
interfaces between payload and spacecraft bus elements.  For both
satellites, the payload module is clearly identified and assembled
inside the structural 1194\,mm central tube. In addition, a large
number of satellite building blocks can be common to the two
satellites, in order to ease mission procurement and tests. This
configuration is particularly compatible with the ESA procurement
scheme.  The payload is made of 3 subsystems: primary mirror and its
dynamic support, the focal plane with its detectors and the
metrology. 

\textbf{Alternative mission concept}.
\begin{figure}[t]
  \centering
  \includegraphics[width=\hsize]{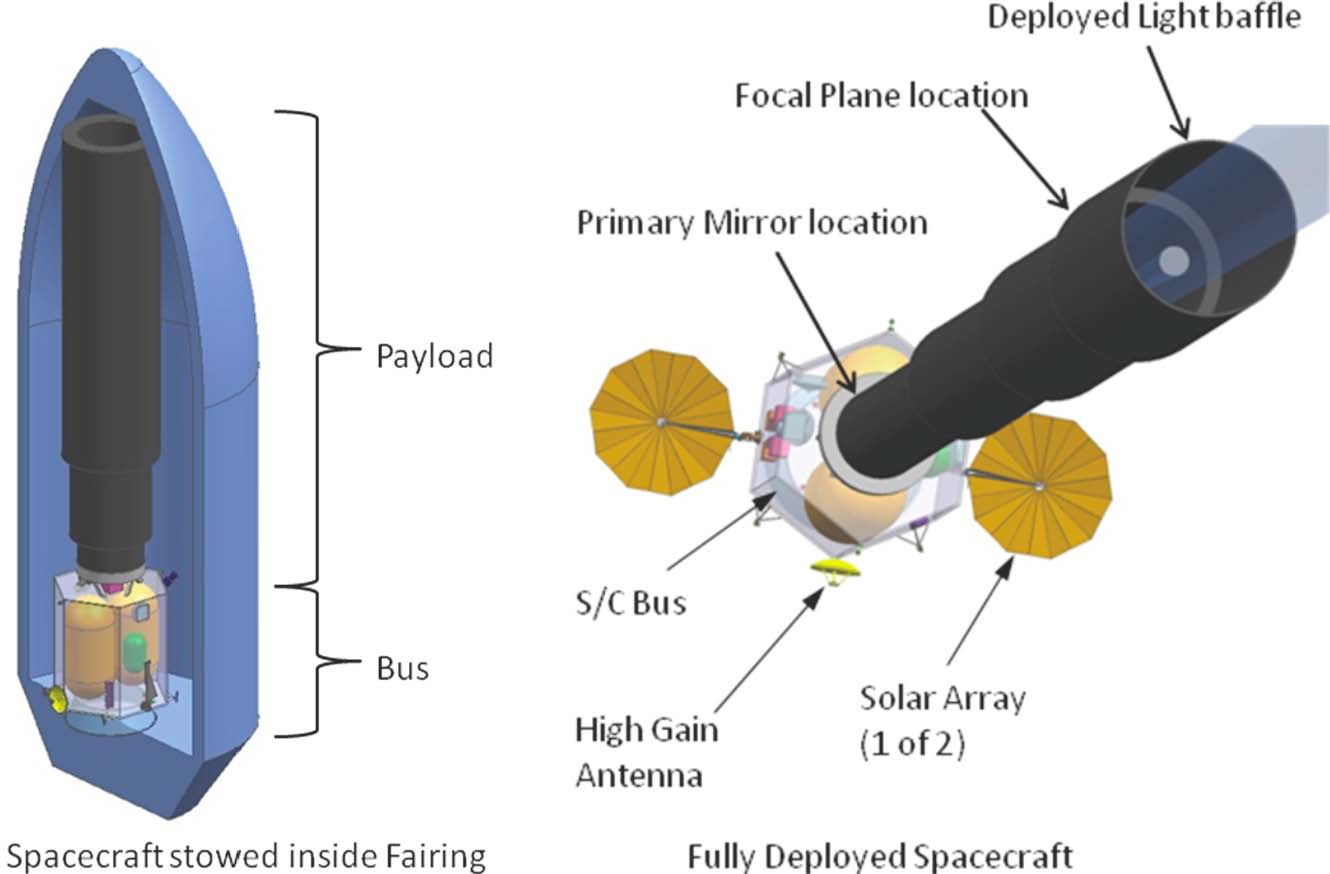}
  \caption{Flight system concept for the deployable telescopic tube version}
  \label{fig:alt-concept}
\end{figure}
An alternative mission concept would consist of a single spacecraft
with an ADAM-like\footnote{ADAM: ABLE Deployable Articulated Mast}
deployable boom (from ATK-Able engineering) that connects the
telescope and the focal plane modules. The preliminary investigation
made by CNES identified no show-stoppers for this option: no
prohibitive oscillation modes during observation; during maneuvers,
the boom oscillation modes can be excited but they can be filtered by
Kalman filters (SRTM\footnote{Shuttle Radar Topography Mission}
demonstration). The use of dampers on the boom structure allows
damping at a level of 10\% of the oscillations. The main worry
concerns retargeting, which requires large reaction wheels or control
momentum gyroscopes (CMGs) on the spacecraft due to the important
inertia but propellers could be added at the boom end. A possible
implementation made by JPL is shown in Fig.~\ref{fig:alt-concept}.

\section{Discussion}
\label{sec:discussion}

\subsection{Astrophysical issues}
\label{sec:astro-challenges}

\textbf{Stellar activity.} If all instrumental problems are controlled
then the next obstacle to achieve the scientific objective is of
astrophysical nature, the impact of stellar activity. Spots and bright
structures on the stellar surface induce astrometric, photometric and
RV signals. Using the Sun as a proxy, \citet{2011A&A...528L...9L} have
computed the astrometric, photometric and RV variations that would be
measured from an observer located 10 pc away. It appears that the
astrometric variations due to spots and bright structures are small
compared to the signal of an Earth mass planet in the HZ \citep[see
also][]{2010A&A...519A..66M, 2009ApJ...707L..73M,
  2010ApJ...717.1202M}. This remains true throughout the entire solar
cycle. If we consider a star \emph{5 times more active than the active
  Sun}, an Earth-mass planet would still be detectable even during the
highest activity phases. Such activity, or lower, translates in terms
of activity index $\log(R'_{HK}) \leq -4.35$. Consequently, in our
target list, we have kept only stars with such an index (\emph{only
  4\% were discarded}), for which their intrinsic activity should not
prevent the detection of an Earth-mass planet, even during its high
activity period.

\textbf{Perturbations from reference stars.} The vast majority of the
reference stars will be K giants at a distance of $\approx
1$\,kpc. The important parameters in addition to the position are the
proper motion with typical value of $\approx 1$\,mas/yr and the
parallax whose typical value is $\approx 1$\,mas.  They are to be
compared to the accuracy of the cumulative measurements during a
visit. An important value for NEAT accuracy is what is obtained for an
$R=6$ magnitude target: 0.8\,\uas/h. The ratios between that
(required) accuracy and the expected motions of the references
indicates clearly that the latter cannot be considered as fixed. Their
positions are members of the set of parameters that have to be solved
for. Because the reference stars are much more distant
($\approx1$\,kpc) than the target star ($\approx10$\,pc), we are 100
times less sensitive to their planetary perturbations. Only
Saturn-Jupiter mass objects matter, and statistically, they are only
present around $\approx10$\% of stars. These massive planets can be
searched for by fitting first the reference star system ($\approx 100
N_{\rm ref}$ measurements for $5 N_{\rm ref}$ parameters when there
are no giant planets around the reference stars), possibly eliminate
those with giant planets, and studying the target star with respect to
that new reference frame. Moreover, the largest disturbers will be
detected from ground based radial velocity measurements, and the early
release of Gaia data around 2016 will greatly improve the position
accuracy of the reference stars. For smaller planets at or below the
threshold of detection, their impact on the target astrometry will be
only at a level $\ll 1\,\MEarth$ around it. Similarly the activity of
these K giants has been investigated and neither the stellar
pulsations nor the stellar spots will disturb the signal at the
expected accuracy.

\textbf{Planetary system extraction from astrometric data.} We
recently carried out a major numerical simulation to test how well a
space astrometry mission could detect planets in multi-planet systems
\citep{2010EAS....42..191T}.  The simulation engaged 5 teams of
theorists who generated model systems, and 5 teams of double-blind
\emph{``observers''} who analyzed the simulated data with noise
included.  The parameters of the study were the same as for NEAT,
viz., astrometric single-measurement uncertainty (0.80\,\uas noise,
0.05\,\uas floor, 5-year mission, plus RV observations with 1\,m/s
accuracy for 15\,years).  We found that terrestrial-mass,
habitable-zone planets ($\approx$ Earths) were detected with about the
same efficiency whether they were alone in the system or if there were
several other giant-mass, long-period planets ($\approx$ Jupiters)
present. The reason for this result is that signals with unique
frequencies are well separated from each other, with little
cross-talk.  The number of planets per system ranged from 1 to 11,
with a median of 3. The SNR value of 5.8 value was predicted by
\citet{1982ApJ...263..835S} for a false alarm probability (FAP) of
less than 1\%, and verified in our simulations. The completeness and
reliability to detect planets was better than 90\% for all planets,
where the comparison is with those planets that should have been
detected according to a Cramer-Rao estimate
\citep{2010ApJ...720.1073G} of the mission noise.  The Cramer-Rao
estimates of uncertainty in the parameters of mass, semi-major axis,
inclination, and eccentricity were consistent with the “observed”
estimates of each: 3\% for planet mass, $\approx4^{\circ}$ for 
inclination and 0.02 for eccentricity.

\textbf{Radial velocity screening.} To solve
unambiguously for giant planets with periods longer than 5\,yrs, it is
necessary to have a ground RV survey for 15\,yrs of the 200 selected
target star, at the presently available accuracy of 1\,m/s. More than
80\% of our targets are already being observed by RV, but the
observations of the rest of them should start soon, well before the
whole NEAT data is available.  The capability of ground based RV
surveys, despite their impressive near-term potential to obtain
accuracies better than 1\,m/s, is not sufficient to detect terrestrial
planets in the HZ of F, G and K stars.  Formally, an accuracy
of 0.05\,m/s is required to see an edge-on Earth mass planet at 1\,\AU
from a solar-mass star with SNR=5 (semi-amplitude = 0.13\,m/s), which
might be achievable instrumentally, but is stopped in most cases by
the impact of stellar activity on RV accuracy. It is necessary to find
particularly ``quiet'' stars, but they are a minority (few percents)
and cannot provide a full sample. Furthermore, the ambiguity in
physical mass associated with the signal coming only from the radial
component of the stellar reflex motion ($\sin i$ ambiguity) requires
additional information to determine the physical mass and relative
inclination in complex planetary systems. In some, but not all cases,
limits are possible, and one can argue statistically that 90\% of
systems should be oriented such that the physical planet mass is
within a factor of two of the mass found in RV. However, for finding a
small number of potential future targets for direct detection and
spectroscopy, an absolute determination that the mass is Earth-like is
required as well as an exhaustive inventory of the planets around
stars in our neighborhood.

\begin{table*}[t]
  \centering
 \caption{Science impact of NEAT scaling. The nominal mission is
    highlighted in yellow.}
  \label{tab:scaling}
  \includegraphics[trim=0mm 7mm 5mm 5mm,clip, width=0.8\textwidth]{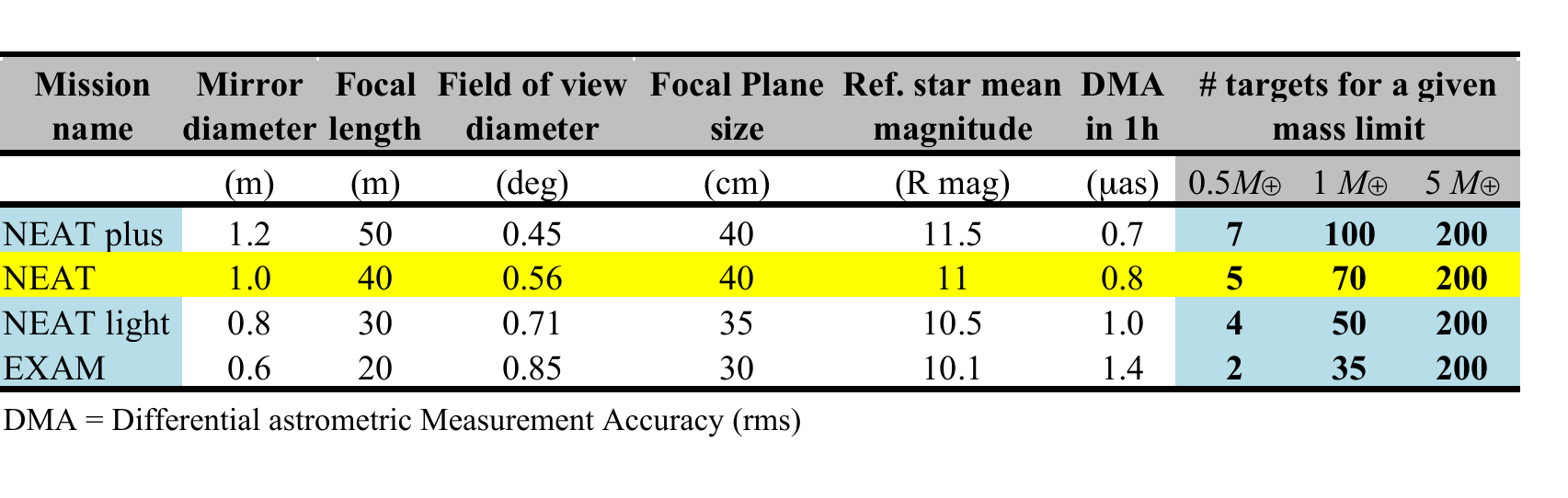}
\end{table*}
\textbf{Flexibility of objectives to upgrades / downgrades of the
  mission.} One of the strengths of NEAT is its flexibility, the
possibility to adjust the size of the instrument with impacts on the
science that are not prohibitive. The size of the NEAT mission could
be reduced (or increased) with a direct impact on the accessible
number of targets but not in an abrupt way. For instance, for same
amount of integration time and number of maneuvers, the options listed
in Table \ref{tab:scaling} are possible, with impacts on the number of
stars that can be investigated down to 0.5 and 1 Earth mass, and on
the mass of the instrument, required fuel for maneuvers, and therefore
cost. The time necessary to achieve a given precision depends on the
mass limit that we want to reach: going from 0.5\,\MEarth to
1\,\MEarth requires twice less precision and therefore 4 times less
observing time allowing a smaller telescope. There is room for
adjustment keeping in mind that one wants to survey the neighborhood
with the smallest mass limit possible and a typical number of targets
of $\approx 200$.

\subsection{Technical issues}
\label{sec:technical-challenges}

\textbf{Optical aberrations.} NEAT uses a very simple telescope
optical design. A 1-m diameter clear aperture off-axis parabola, with
an off-axis distance of 1\,m and a 40\,m focal length. The focal plane
is at the prime focus. The telescope is diffraction limited at the
center of the field, where the target stars will be observed, but coma
produces some field dependent aberrations. At the mean position of the
reference stars, $0.2^{\circ}$ away from the center of the field, the
coma produces a \emph{steady} 23\% increase of the point spread
function (PSF) width and an 8\,\microns centroid offset. The impact
remains low since we are looking at differential effects.

\textbf{Centroid measurements.} They consist of two steps: the
determination of the stellar centroid on each CCD during 57\,s and
then the calibration of the relative position of the CCDs during 3\,s
thanks to the metrology. The metrology determines also the response
map of the detectors. As in the normal approach to precision
astrometry with CCDs, we perform a least-square fit of a template PSF
to the pixelated data. PSF knowledge error leads to systematic errors
in the conventional centroid estimation. We have developed an accurate
centroid estimation algorithm by reconstructing the PSF from well
sampled (above Nyquist frequency) pixelated images. In the limit of an
ideal focal plane array whose pixels have identical response function
(no inter-pixel variation), this method can estimate centroid
displacement between two 32x32 images to sub-micropixel
accuracy. Inter-pixel response variations exist in real CCDs, which we
calibrate by measuring the pixel response of each pixel in Fourier
space\footnote{They are determined by calculating the first 6
  coefficients of the Taylor series expansion in powers of wave
  numbers of the detector response map Fourier components.}. Capturing
inter-pixel variations of pixel response to the third order terms in
the power series expansion, we have shown with simulated data that the
centroid displacement estimation is accurate to a few micro-pixels.

\textbf{Stability of the primary mirror.} The primary optic will be
made of zerodur/ULE with a temperature coefficient better than
$10^{-8}$/K with an optics thickness $\approx 10$\,cm and
the effective temperature and temperature gradients are kept stable to
$\approx 0.1$\,K over the mirror, the optic is then stable to
$\approx0.1$\,nm ($\lambda/6000$) during the 5\,yr
mission. We have simulated two images, one at the center of the field
that is a perfect Airy function and one at the edge of the field that
has a $\lambda/20$ coma. We added also wavefront errors with a
conservative rms value of $\lambda/1000$. With the new wavefronts, we
calculated the change in the differential astrometry bias caused by
both pixelation and changing wavefronts. While the wavefront
deviations to optimal shape caused a centroid shift of
$\approx6-10\,\uas$ ($10^{-4}$ pixels), differential errors remained
less than $\approx0.3\,\uas$ ($3\times10^{-6}$ pixels).

\textbf{CCD damage in L2 environment.} CCDs, like most semiconductors,
suffer damage in radiation environments such as encountered by space
missions. One particular performance parameter, Charge Transfer
Efficiency (CTE), degrades with known consequences on the efficiency
of science missions like Gaia\footnote{The Gaia community
  (\url{http://www.rssd.esa.int/gaia}) speaks of the complementary
  quantity, charge transfer inefficiency (CTI), in order to emphasize
  its detrimental effects.}. The reduced CTE is caused generally by
prompt particle events (PPE), including solar protons and cosmic rays,
colliding with the CCD silicon lattice and causing damages to the
silicon lattice. This leads to the formation of so-called traps which
can capture photo-electrons and release them again after some
time. This results in signal loss and distortion of the PSF shape. The
latter leads to systematic errors in the image location due to a
mismatch between the ideal PSF shape and the actual image shape. For
Gaia, this effect of radiation damage is a major contributor to the
error budget and extensive research and laboratory tests have been
done in order to understand better the radiation damage effects and to
develop approaches in both hardware and data processing to mitigate
the negative impact. However, there are a number of important
differences between NEAT and Gaia which justify the assumption that
radiation damage effects will play a much smaller role: i) NEAT looks
for extended periods at very bright stars compared to Gaia in which
the stars continuously move on the CCD. Also, unlike Gaia, NEAT will
not be operated in time-delayed integration mode. In addition the CCDs
are regularly illuminated by the laser light from the metrology
system. This means that in general the signal level in the CCD pixels
is high which will keep the traps with long ($\geq60$\,s) release time
constants filled and effectively inactive. ii) NEAT also does not
suffer from the varying CCD illumination history that a scanning
mission like Gaia necessarily encounters. This illumination history is
in fact one of the major complicating factors for Gaia. Finally, iii)
NEAT uses much smaller CCDs than Gaia and in addition has four
read-out nodes, thus reducing the number of charge transfer steps and
mitigating the effects of radiation damage. The one concern for the
NEAT case is the presence of traps with release time constants that
are of the order of several times the charge transfer period between
pixels. In the case of NEAT the transfer period averages tens of
$\mu$s and from laboratory tests with E2V CCDs, carried out in the
context of the Gaia project, traps with time constants of
10--100\,$\mu$s are known to exist. If these traps dominate at the
operating temperature of the NEAT CCDs they could lead to subtle PSF
image shape distortions and thus image location biases. From the Gaia
experience, it is known that such image shape distortions can be
handled in the post-processing by a careful modeling of the effects of
radiation damage on the PSF image. A similar strategy, building on the
Gaia heritage, can be employed for NEAT.

\begin{figure*}[t]
  \centering
  \includegraphics[width=\textwidth]{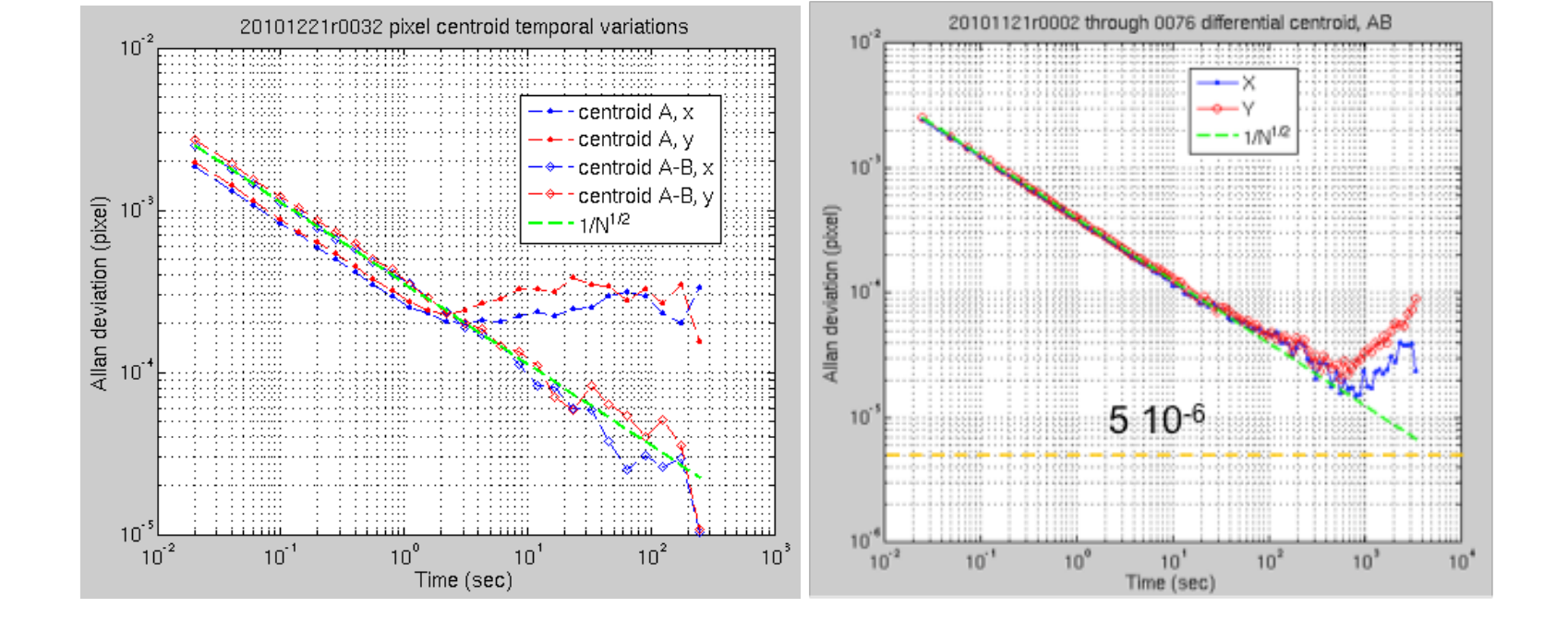}
  \caption{Latest results obtained from the CCD/metrology test
    bench. No metrology nor quantum efficiency 6-parameter calibration
    have been performed yet. Left: Allen deviation of the centroid
    location for one artificial star ``A'' projected on the CCD and
    Allen deviation of the differential centroid location for two
    artificial stars A and B projected on the CCD. One can see that
    star A moves on the CCD by a few hundred micro-pixels at time
    scale greater than 1 second, but that the differential position of
    the two stars is better $4\,10^{-5}$ pixel at 100\,s integration
    time. Right: Allen deviation of the differential centroid location
    for two artificial stars projected on the CCD. Concatenating data
    from 38 runs, a minimum of $\approx 20\,\mu$pixels at about 10\,min before
    differential drift dominated.  This data shows that differential
    metrology at intervals of minutes is needed.}
  \label{fig:testbed-results}
\end{figure*}
\textbf{CCD/metrology tests in the lab.} In the absence of optical
errors, the major error sources are associated with the focal plane:
(1) motions of the CCD pixels, which have to be monitored to
$3\times10^{-6}$ pixels every 60\,s, i.e.\ 0.03\,nm; (2) measurements
of the centroid of the star images with $5\times10^{-6}$ pixel
accuracy. We have set up technology testbeds to demonstrate that we
can achieve these objectives. The technology objective for (1) has
almost been reached and the technology demonstration for (2) is
underway and should be completed soon. Latest results with no
metrology nor QE 6-parameter calibration have been obtained from the
CCD / metrology test bench (Fig.~\ref{fig:testbed-results}). Allen
deviation of the centroid location for one artificial star ``A''
projected on the CCD and Allen deviation of the differential centroid
location for two artificial stars A and B projected on the CCD are
plotted in Fig.~\ref{fig:testbed-results} (left). One can see that
star A moves on the CCD by a few hundred micro-pixels at time scale
greater than 1 second, but that the differential position of the two
stars is better $4\times10^{-5}$ pixel at 100s integration time. On
Fig.~\ref{fig:testbed-results} (right), the Allen deviation of the
differential centroid location for two artificial stars projected on
the CCD is plotted, concatenating data from 38 runs, a minimum of
$\approx20\,\mu$-pixels at about 10\,min before differential drift
dominated. This data shows that we are only a factor 10 from the final
goal and that differential metrology at intervals of minutes is
required to reach it.

\section{Perspectives}
\label{sec:perspectives}

In the Cosmic Vision plan for 2015-2025, the community has identified
in Theme 1 the question: \emph{``What are the conditions for planet
  formation?''}, and the recommendation in Sect.~1.2: \emph{``Search
  for planets around stars other than the Sun...''} ultra high precise
astrometry as a key technique to explore our solar-like neighbors.
\begin{quote} \em
  ``On a longer timescale, a complete census of all Earth-sized
  planets within 100 pc of the Sun would be highly desirable. Building
  on Gaia's expected contribution on larger planets, this could be
  achieved with a high-precision terrestrial planet astrometric
  surveyor.''
\end{quote}
We have designed NEAT to be this astrometric surveyor. In Europe, as
discussed in detail in the conclusions of the conference
\emph{Pathways to Habitable Planets} \citep{2010ASPC..430.....C} and
in the \emph{Blue Dot Team} report, the exoplanet community recognizes
the importance of astrometric searches for terrestrial planets and has
prioritized this search as a key question in the mid-term, i.e. in the
time frame 2015-2022. The ExoPlanet Task Force (ExoPTF) in the US made
a similar statement. Finally the ESA dedicated \emph{ExoPlanetary
  Roadmap Advisory Team} (EPRAT) prioritizes \emph{Astrometric
  Searches for Terrestrial Planets} in the mid term, i.e.\ in the time
frame 2015-2022. Although the Decadal Survey of Astronomy and Astrophysics for
2010-2020 ranked down the SIM-Lite proposal, but placed as number one
priority a program \emph{``to lay the technical and scientific foundation for
a future mission to study nearby Earth-like planets''}.

Because of these recommandations by the community, we believe that
there is a place for a mission like NEAT in future space programs,
that is to say, a mission that is capable of detecting and
characterizing planetary systems orbiting bright stars in the solar
neighborhood that have a planetary architecture like that of our Solar
System or an alternative planetary system partly composed of
Earth-mass planets. These stars visible with the naked eye or simple
binoculars, if found to host Earth-mass planets, will change
humanity's view of the night sky.

\begin{acknowledgements}
This work has benefited support from the Centre National des \'Etudes
Spatiales (CNES), the Jet Propulsion Laboratory (JPL), Thales Alenia
Space (TAS) and Swedish Space Corporation (SSC).
\end{acknowledgements}



\section*{Author affiliations}

\newcommand{\and}{--~}
\newcommand{\at}{:~}

\begin{list}{}{\setlength{\itemsep}{1em}}
\item   F.~Malbet \and Ph.~Feautrier \and A.-M. Lagrange \and A.~Chelli \and
  G.~Duvert \and T.~Forveille \and N.~Meunier \at UJF-Grenoble 1 /
  CNRS-INSU, Institut de Plan\'etologie et d'Astrophysique de Grenoble
  (IPAG), UMR 5274, BP 53, F-38041 Grenoble cedex 9, France.
  \email{Fabien.Malbet@obs.ujf-grenoble.fr}
\item A.~L\'eger \at Universit\'e Paris Sud / CNRS-INSU, Institut
  d'Astrophysique Spatiale (IAS) UMR 8617, B\^at.~120-121, F-91405
  Orsay cedex, France
\item M.~Shao \and R.~Goullioud \and W.~Traub \at Jet Propulsion
  Laboratory (JPL), California Institute of Technology, 4800 Oak Grove
  Drive, Pasadena CA 91109, USA
\item P.-O.~Lagage \and C.~Cara \and G.~Durand \at Laboratoire AIM,
  CEA-IRFU / CNRS-INSU / Universit\'e Paris Diderot, CEA Saclay, B\^at
  709, F-91191 Gif-sur-Yvette Cedex, France
\item A.G.A.~Brown \and H.J.A.~R\"ottgering \at Leiden Observatory,
  Leiden University, P.O.\ Box 9513, NL-2300 RA Leiden, The
  Netherlands
\item C.~Eiroa \and J.~Maldonado \and E.~Villaver \at Universidad
  Aut\'onoma de Madrid (UAM), Dpto.\ F\'isica Te\'orica, M\'odulo 15,
  Facultad de Ciencias, Campus de Cantoblanco, E-28049 Madrid, Spain
\item B.~Jakobsson \at Swedish Space Corporation (SSC), P.O.\ Box
  4207, SE-17104 Solna, Sweden
\item E.~Hinglais \at Centre National d'Etudes Spatiales (CNES),
  Centre spatial de Toulouse, 18 avenue Edouard Belin, F-31401
  Toulouse cedex 9, France
\item L.~Kaltenegger \and C.~Mordasini \at Max-Planck-Institut f\"ur
  Astronomie, K\"onigstuhl 17, D-69117 Heidelberg, Germany
\item L.~Labadie \at I.\ Physikalisches Institut der Universit\"at zu
  K\"oln, Z\"ulpicher Str.\ 77, D-50937 K\"oln, Germany
\item J.~Laskar \and N.~Rambaux \at UPMC-Paris 6 / Observatoire de
  Paris / CNRS-INSU, Institut de m\'ecanique c\'eleste et de calcul
  des \'eph\'em\'erides (IMCCE) UMR 8028, 77 avenue Denfert-Rochereau,
  F-75014 Paris, France
\item R.~Liseau \at Chalmers University of Technology, SE-41296
  Gothenburg, Sweden
\item J.~Lunine \at Dipartimento di Fisica, University of Rome Tor
  Vergata, Via della Ricerca Scientifica 1, Roma I-00133, Italy
\item M.~Mercier \at Thales Alenia Space, 100 boulevard du Midi,
  F-06150 Cannes, France
\item D.~Queloz \and J.~Sahlmann \and D.~S\'egransan \at Observatoire
  Astronomique de l'Universit\'e de Geneve, 51 Chemin des Maillettes,
  CH-1290 Sauverny, Switzerland
\item A.~Quirrenbach \at Universit\"at Heidelberg, Landessternwarte,
  K\"onigstuhl 12, D-69117 Heidelberg, Germany
\item A.~Sozzetti \and A.H.~Andrei \at INAF - Osservatorio Astronomico
  di Torino, Strada Osservatorio 20, I-10025 Pino Torinese, Italy
\item O.~Absil \and J.~Surdej \at Universit\'e de Li\`ege,
  D\'epartement d'Astrophysique, G\'eophysique et Oc\'eanographie, 17
  All\'ee du Six Ao\^ut, B-4000 Sart Tilman, Belgium
\item Y.~Alibert \at Physikalisches Institut, University of Bern,
  Sidlerstrasse 5, 3012, Bern, Switzerland
\item Y.~Alibert \at Universit\'e de Besan\c{c}on / Observatoire de
  Besan\c{c}on / CNRS-INSU UMR 6213, Institut UTINAM, 41 bis Avenue de
  l'Observatoire, BP 1615, F-25010 Besan\c{c}on cedex, France
\item A.H.~Andrei \at Observatorio Nacional - Minist\'erio da
  Ci\^encia e Tecnologia, Rua General Cristino 77, S\=ao Crist\'ov\=ao
  20921-400, Rio de Janeiro, Brazil
\item F.~Arenou \at Universit\'e Paris 7 Diderot / Observatoire de
  Paris / CNRS-INSU, UMR 8111, "Galaxie – Etoile – Physique –
  Instrumentation" (GEPI), 5 Place Jules Janssen, F-92190 Meudon
  cedex, France
\item C.~Beichman \at California Institute of Technology, NASA
  Exoplanet Science Institute / IPAC, 770 South Wilson Ave, Pasadena
  CA 91125, USA
\item C.S.~Cockell \at The Open University, Dept.\ of Physics \&
  Astronomy, Planetary and Space Sciences Research Institute, Milton
  Keynes MK7 6AA, UK
\item P.J.V.~Garcia \at Universidade do Porto, Faculdade de
  Engenharia, Departamento de Engenharia Fí­sica, Laboratório SIM, Rua
  Dr. Roberto Frias, P-4200-465 Porto, Portugal
\item D.~Hobbs \at Lund Observatory, Lund University, Box 43, SE-22100
  Lund, Sweden
\item A.~Krone-Martins \at Instituto de Astronomia, Geof\'isica e
  Ci\^encias Atmosf\'ericas, Universidade de S\~ao Paulo, Rua do
  Mat\~ao, 1226, Cidade Universit\'aria, 05508-900 S\~ao Paulo-SP,
  Brazil
\item A.~Krone-Martins \and S.~Raymond \and F.~Selsis \at Universit\'e
  de Bordeaux 1 / Observatoire Aquitain des Sciences de l'Univers /
  CNRS-INSU, UMR 5804, Laboratoire d'Astrophysique de Bordeaux, BP 89,
  F-33271 Floirac Cedex, France
\item H.~Lammer \at Space Research Institute, Austrian Academy of
  Sciences, Schmiedlstr.\ 6, A-8042 Graz, Austria
\item S.~Minardi \at Institute of Applied Physics, Friedrich Schiller,
  University Jena, Max-Wien-Platz 1, D-07743 Jena, Germany
\item A.~Moitinho de Almeida \and A.~Krone-Martins \at Systems,
  Instrumentation and Modeling (SIM) - Faculdade de Ci\^encias da
  Universidade de Lisboa, Ed. C8, Campo Grande, 1749-016 Lisboa,
  Portugal
\item P.A.~Schuller \at Universit\'e Paris 7 Diderot / Universit\'e
  Pierre et Marie Curie / Observatoire de Paris / CNRS-INSU, UMR 8109,
  Laboratoire d'\'etudes spatiales et d'instrumentation en
  astrophysique (LESIA), 5 Place Jules Janssen, F-92190 Meudon cedex,
  France
\item G.J.~White \at The Open University, Dept.\ of Physics \&
  Astronomy, Venables Building, Walton Hall, Milton Keynes MK7 6AA, UK
\item G.J.~White \at Rutherford Laboratory, Space Science \&
  Technology Department, CCLRC Rutherford Appleton Laboratory,
  Chilton, Didcot, Oxfordshire OX11 0QX, UK
\item H.~Zinnecker \at Deutsches SOFIA Institut, Institut f\"ur
  Raumfahrtsysteme, Universit\"at Stuttgart, Pfaffenwaldring 31,
  D-70569 Stuttgart, Germany
\item H.~Zinnecker \at SOFIA Science Center, NASA-Ames, MS 211-3,Moffett Field, CA 94035, USA
\end{list}
\end{document}